\documentclass[prd,floats,superscriptaddress,nofootinbib,preprintnumbers,twocolumn]{revtex4}%
\usepackage{lipsum, babel}

\usepackage{amssymb}
\usepackage{amsmath}

\usepackage{sans}

\makeatletter\AtBeginDocument{\let\@elt\relax}\makeatother

\usepackage{graphicx}
\usepackage{graphics}
\usepackage{dcolumn}
\usepackage{color}
\usepackage{rotate}
\usepackage{fancyhdr} 
\usepackage{hyperref}
\usepackage{indentfirst}
\usepackage{booktabs} % LT for tables
\usepackage{fontawesome} % LT for github logo
\usepackage{physics} % LT for useful routines
\usepackage{siunitx} % For SI stuff
\DeclareSIUnit \parsec {pc}
\DeclareSIUnit \year {yr}
\usepackage{umoline}
\usepackage{enumitem}

\def\be{\begin{eqnarray}}
\def\ee{\end{eqnarray}}
\def\ba{\begin{eqnarray}}
\def\ea{\end{eqnarray}}

\definecolor{darkred}{rgb}{.743,0,0}

\newcommand{\iu}{\mathrm{i}}
\newcommand{\e}{\mathrm{e}}
\begin{document}
\title{
AxionH0graphy: hunting for ultralight dark matter with cosmographic H$_0$ signal}

\author{Kfir Blum}\email{kfir.blum@weizmann.ac.il}
\affiliation{Weizmann Institute of Science, Rehovot 7610001, Israel} 
 \author{Luca Teodori}\email{luca.teodori@weizmann.ac.il}  
 \affiliation{Weizmann Institute of Science, Rehovot 7610001, Israel} 
%\date{\today}

\begin{abstract}
If ultralight boson fields exist, 
then vacuum misalignment populates them with nonzero relic abundance. For a broad range of particle mass $m$ the field condenses into fuzzy cores in massive galaxies. 
We use numerical simulations to test this idea, extending previous work (Blum \& Teodori 2021~\cite{Blum:2021oxj}) and focusing on ultralight dark matter (ULDM) that makes-up a subdominant fraction of the total dark matter density, consistent with observational constraints. Our simulations mimic galactic halos and explore different initial conditions and levels of sophistication in the modeling of the halo potential. For $m\sim10^{-25}$~eV ULDM cores act as approximate internal mass sheets in strong gravitational lensing, and could first be detected using time-delays in cosmography, when an $H_0$ prior is assumed: a scenario we dub AxionH0graphy. 
The mass sheet degeneracy is broken by finite core radius and by the dynamical displacement of cores from the halo center of mass, that introduce imaging distortions and restrict the $H_0$ signal limit of AxionH0graphy to $m\lesssim5\times10^{-25}$~eV. 
Cosmological simulations are called for to sharpen the predicted connection between the amplitude of ULDM galactic cores and the ULDM cosmological fraction. 
\end{abstract}
%\pacs{}

\maketitle

\tableofcontents
%%%%%%%%%%%%%%%%%%%
\section{Introduction}
UltraLight Dark Matter (ULDM) could arise via vacuum misalignment~\cite{Preskill:1982cy,Abbott:1982af,Dine:1982ah}, and the scenario of particle mass as low as $ m < 10^{-20} $ eV has been discussed extensively in the literature~\cite{Hu:2000ke,Hui2017,Hui:2021tkt,Ferreira:2020fam}. A plethora of axion-like particles spanning a range of masses may occur in string theory~\cite{Svrcek:2006yi,Arvanitaki:2009fg,Marsh:2015xka}, motivating the possibility of several ULDM fields, each contributing a fraction of the DM budget.  

ULDM differs from collisionless Cold DM (CDM) in several features, allowing to constrain the ULDM mass-fraction using gravitational probes alone, without relying on model-dependent assumptions. From a cosmological point of view, ULDM suppresses power on small scales, making it possible to constrain it via Cosmic Microwave Background (CMB) anisotropies and galaxy clustering~\cite{Lague:2021frh}, as well as the Lyman-alpha forest~\cite{Irsic:2017yje,Armengaud:2017nkf,Kobayashi:2017jcf,Leong:2018opi}. 
In galaxies, ULDM interference patterns characterized by the de Broglie wavelength $\lambda_{\rm dB}\approx 2\pi/(mv)$, where $v$ is the characteristic velocity in the system, are seen in simulations~\cite{Guzman2004,Schive:2014dra,Schive:2014hza,Schwabe:2016rze,Veltmaat:2016rxo,Mocz:2017wlg,Veltmaat:2018dfz,Levkov:2018kau,Eggemeier:2019jsu,Chen:2020cef,Schwabe:2020eac,Zhang:2018ghp}, and lead to detectable effects related to dynamical heating and dynamical friction~\cite{Hui2017,Amorisco:2018dcn,Church:2018sro,Lancaster:2019mde,Bar:2021jff,DuttaChowdhury:2023qxg}. 
Another prominent feature of ULDM in galaxies is the formation of cored profiles in the inner part of the dark matter halo~\cite{Schive:2014dra,Schive:2014hza,Hui2017}. These cores, called ``solitons" in the ULDM literature, are the ground state solution of the Schr\"odinger-Poisson equations, obeyed by ULDM in the non-relativistic regime. They are consistently seen in  simulations, and some of their key features are understood theoretically~\cite{Chavanis:2011zi,Chavanis:2011zm,Levkov:2018kau,Guth:2014hsa,Bar-Or:2018pxz,Bar-Or:2020tys,Chan:2022bkz,Chan:2023crj}. For $m\lesssim10^{-21}$~eV, the ULDM core prediction is in tension with observations of the rotation curves of low-surface-brightness galaxies, if a single ULDM species makes up the full DM budget~\cite{Bar:2018acw,Bar:2021kti,Bernal:2017oih}. 

Gravitational lensing offers another probe of ULDM. Magnification anomalies can be used to constrain the amount of substructure present in a galaxy~\cite{Koopmans:2005nr,Xu:2013kna,Vegetti:2008eg,Vegetti:2023mgp}, including substructure due to ULDM~\cite{Laroche:2022pjm,Powell:2023jns}. 
Here we focus on a different effect, suggested in Ref.~\cite{Blum:2021oxj}, 
that concerns the impact of ULDM cores on quasar time delays in strong gravitational lensing systems. Such cores affect the measurement of the Hubble parameter $ H_0 $ inferred from time delay cosmography~\cite{Millon:2019slk}, acting as an approximate internal Mass-Sheet Degeneracy (MSD)~\cite{Falco1985,Schneider:2013sxa,Blum:2020mgu}.
%\footnote{An internal MSD effect could also come from mis-modeling the host group of the lens.}
%\footnote{Lens systems with multiple sources (e.g.~\cite{DES:2019fny,Ballard:2023fgi}) may provide another avenue to detect or constrain ULDM cores. Although the MSD is formally not broken in multi-source systems~\cite{McCully:2013fga,Schneider:2014vka,Schneider:2014ifa}, physical modeling assumptions~\cite{Teodori:2022ltt} could point to internal MSD-like effect in the main lens via the inference of angular diameter distance ratios. Many multi-source systems are expected to be discovered in the near future~\cite{Oguri:2010ns, Gavazzi:2008aq,Liao:2014cka}.}  

Some studies accounted for the MSD by direct implementation in the analysis pipeline, either via an effective ideal mass sheet~\cite{Birrer:2020tax}, 
% (which, however, may introduce bias when coupled to stellar kinematics data~\cite{Blum:2021oxj}), 
or via a particular core parametrization~\cite{Shajib:2023uig}. Such analyses are timely both because they could shed light on the $ H_0 $ tension~\cite{Verde:2019ivm,DiValentino:2021izs,Freedman:2023jcz,Verde:2023lmm,Blum:2020mgu}, and because current and upcoming surveys~\cite{DES:2018whv,SKA:2018ckk,Euclid:2019clj,LSSTScience:2009jmu} are expected to discover many new time-delay systems~\cite{Oguri:2010ns,Dobke:2009bz, Gavazzi:2008aq,Liao:2014cka}, possibly allowing a $ < $2\% measurement of $ H_0 $ with improved stellar kinematics~\cite{Birrer:2020jyr}. 

Interestingly, Ref.~\cite{Birrer:2020tax} found some positive preference (at the $2\sigma$ level) for an internal mass sheet component, when combining time-delays and lens morphology prior based on a larger galaxy sample. 
It is amusing to entertain the possibility that this could be the first observational evidence of ULDM and thus, perhaps, a hint of ultra-high energy physics. More generally, the results of~\cite{Birrer:2020tax,Blum:2021oxj} suggest that cosmography is sensitive to features in massive galaxy morphology on 10-100~kpc scales, that are difficult to detect via kinematics, weak lensing, or large-scale cosmological data. As noted in~\cite{Blum:2020mgu}, this sensitivity can be significantly improved if one includes external prior constraints on the value of $H_0$, coming from other data sets.

In this paper we continue the exploration of ULDM cores in relation to gravitational lensing. %However, all these prospects require understanding on how these cores form, and how massive are expected to be.  
Ref.~\cite{Blum:2021oxj} showed that, for ULDM cores to be diffuse enough to behave as approximate mass sheets, to be compatible with MSD constraints from stellar kinematics, and massive enough to yield detectable $H_0$ bias at or exceeding a few percent, requires $10^{-25}\lesssim  m\lesssim10^{-24}$~eV. Such a bias will become a signal, if an $H_0$ prior, together with explicit core modelling, are used. Employing a lens model with an explicit core component, using imaging data, time delays with an $H_0$ prior, and stellar kinematics measurements (which are also sensitive to a MSD component) can yield a detection of such a core. This is what we call the AxionH0graphy scenario. 

In the range of particle mass relevant for AxionH0graphy, ULDM is constrained to comprise only a fraction $ \alpha_\chi\lesssim 0.2 $ of the total DM budget~\cite{Kobayashi:2017jcf,Lague:2021frh}. 
For such small $\alpha_\chi$, simple estimates of the dynamical relaxation time scale~\cite{Bar-Or:2018pxz,Bar-Or:2020tys} led Ref.~\cite{Blum:2021oxj} to identify the time required to form a core as a bottleneck for the lensing effect. 

We extend the analysis by using numerical simulations %\footnote{Github repository with the code:~\href{https://github.com/lucateo/ULDM_grid}{\faGithub}} 
to study ULDM dynamics on the relevant galactic scales. 
We focus on $10^{-25}\lesssim  m\lesssim10^{-24}$~eV, $\alpha_\chi\approx0.15$ consistent with all other observational constraints, and TDCOSMO-like galaxies~\cite{Millon:2019slk}: massive ellipticals with $ M_{200} \approx 10^{13} M_\odot $ and lens and source redshift configurations typical in strong lensing. 

Our simulations start from variations around a simple class of initial conditions, where the ULDM roughly follows the same distribution as the CDM component. For these initial conditions, the ULDM in the central region of the halo evolves due to wave pressure to form a flattened core of size $\sim\lambda_{\rm dB}$. The system first expels excess mass from the core region (thus mostly erasing the  dependence on the details of the initial conditions in the inner halo). Subsequently, the core density oscillates around a mean level that grows in qualitative agreement with previous related analyses~\cite{Levkov:2018kau}. 

For most simulated cases, the ULDM spatial distribution for $m\lesssim\SI{5e-25}{\electronvolt}$ is consistent with approximate internal MSD for galaxy lensing applications. We believe that this aspect of the results is robust to variations in initial conditions, and to the detailed modeling of the dominant CDM and stellar mass components of the halo. The equivalent $H_0$ signal in our simulations ranges between about $1$ and $20$ percent. For example, for $m=2\times10^{-25}$~eV (still with $\alpha_\chi=0.15$) the mean fractional bias across 6 simulations is $\delta_H=\delta H_0/H_0\approx+0.1$ with standard deviation of roughly $0.06$. Again, such a bias would translate in detection of a core component with convergence $\kappa_{\rm c} \sim 0.1$ inside the lens system under consideration, if an $H_0$ prior is used. In most cases, our estimates suggest that MSD-breaking imaging distortions would be difficult to detect, even without considering degeneracies with other systematic effects like cosmological external shear, sub-halos or host group association. The magnitude of the expected $\kappa_{\rm c}$ could, however, depend on the details of the initial conditions at the outskirts of the halo and may be different for cosmological initial conditions, which we are not yet able to perform in this work. 

The paper is structured as follows. We describe the approximate MSD in gravitational lensing and the relation to ULDM in Sec.~\ref{s:cores}. In Sec.~\ref{s:setup} we explain the simulations. The main results are presented in Sec.~\ref{s:results}, with additional discussion regarding the formation and growth of cores given in Sec.~\ref{s:t}. We summarize in Sec.~\ref{s:sum}. 

App.~\ref{s:code} provides more details of the numerical code. App.~\ref{s:shear} contains a discussion of shear and flexion terms. App.~\ref{s:burk} contains simulation results where ULDM halos are initialized differently, to further explore possible initial conditions effects.
%~\href{https://github.com/lucateo/ULDM_grid}{\faGithub}.
%

\section{ULDM cores and cosmographic $H_0$ signal} \label{s:cores}
Consider a gravitational lensing system characterized by the lens convergence $ \kappa_0(\vec \theta) $, with Einstein angle $ \theta_{\rm E} $. The equation relating the angular position of the source $ \vec{\beta}_0 $ with the observed image angle $ \vec{\theta} $ reads
\begin{equation} \label{eq:infer_model}
\vec{\beta}_0 = \vec{\theta} - \vec{\alpha}_0 (\vec{\theta}) \ ,
\end{equation}
where $ \vec{\alpha}_0 $ is the displacement angle, related to $ \kappa_0 $ via $2\kappa_0(\vec \theta)=\grad_\theta\cdot\vec\alpha_0(\vec\theta)$ (and satisfying $\alpha_0(\theta_{\rm E})=\theta_{\rm E}$ for a spherically-symmetric system). Changing the model parameters via
\begin{equation} \label{eq:msd}
\kappa(\vec{\theta}) = (1- \kappa_{\rm c}(\theta_{\rm E}))\kappa_0(\vec{\theta}) + \kappa_{\rm c}(\vec\theta) \ , 
%\ \kappa_{\rm c} = (1-\lambda)\gamma(\vec{\theta}) \ , 
\ \vec\beta = (1- \kappa_{\rm c}(\theta_{\rm E}))\vec\beta_0\,, 
\end{equation}
keeps imaging data approximately invariant, and acts as an approximate MSD, provided that $\kappa_{\rm c}(\theta)$ is a sufficiently slow-varying function of $\theta$ for $\theta$ smaller than a few times $\theta_{\rm E}$. (In particular, Eq.~(\ref{eq:msd}) expresses an exact MSD if $\kappa_{\rm c}=$~Const.)

Following Ref.~\cite{Blum:2020mgu}, we consider the case that $\kappa_{\rm c}$ arises from a cored mass component associated with the main lens, and call $\kappa_{\rm c}$ a core-MSD component. Following Ref.~\cite{Blum:2021oxj}, the physical mechanism we consider to produce the core is ULDM.

{\it$H_0$ signal:}
Lensing time delays are not invariant under the MSD. If the model of Eq.~\eqref{eq:msd} is the truth model corresponding to the actual physical distribution of mass along the line of sight, while Eq.~\eqref{eq:infer_model} is the model assumed by an inference analysis pipeline, one finds
\begin{equation}\label{eq:dt_bias}
	\frac{\Delta t^{\rm truth} - \Delta t^{\rm inferred}}{\Delta t^{\rm inferred}} \approx \kappa_{\rm c}(\theta_{\rm E}).
\end{equation}
In cosmography, this time delay bias translates into an $ H_0 $ inference bias 
\begin{equation} \label{eq:H0_bias}
	\delta_{H}=%\frac{\delta H_0}{H_0}=
	\frac{H_0^{\rm inferred} - H_0^{\rm truth}  }{H_0^{\rm inferred}} \approx \kappa_{\rm c}(\theta_{\rm E}) \ .
\end{equation}
We assume small $\kappa_{\rm c}(\theta_{\rm E})\ll1$ (our main discussion revolves around $\kappa_{\rm c}(\theta_{\rm E})$ of order 0.1).
Such an $H_0$ bias would transform in a $H_0$ signal of the core component $ \kappa_{\rm c}(\theta_{\rm E})$, if the (supposed) truth model and $H_0$ prior are used.
% 
%It follows that a physical density core component that is not accounted for in the analysis pipeline would bias the inferred value of $H_0$ positively, that is, above the truth value of $H_0$. Conversely, 
An external prior on $H_0$ can allow cosmography analyses to detect density cores~\cite{Blum:2020mgu} down to fractional column densities of order ten percent, that are difficult to detect from kinematics~\cite{Blum:2021oxj}. With this understanding in mind, we will call $\delta_{H}$ the $H_0$ signal from now on. 

Eq.~(\ref{eq:H0_bias}) equates the $H_0$ signal to the core convergence at the Einstein angle. This may seem confusing, because $\kappa_{\rm c}(\theta_{\rm E})$ is the ULDM convergence, not the fractional contribution of ULDM to the total convergence. In practice, however, the $H_0$ signal is indeed proportional to fractional, rather than absolute, ULDM convergence: this is because the observable effect comes from the angular region containing imaging data, which is concentrated around the effective Einstein angle $\theta_{\rm E}$. Focusing on spherical systems for simplicity, the Einstein angle of the null model is given by the solution of 
\be 2\int_0^1dy\,y\,\kappa_0(y\,\theta_{\rm E})&=&1.\ee
The integral is typically dominated by the region $y=\mathcal{O}(1)$, implying $\kappa_0(\theta_{\rm E})=\mathcal{O}(1)$. For example, 3D density $\rho_0(r)\propto r^{-\gamma}$ leads to $\kappa_0(\theta_{\rm E})=(3-\gamma)/2$, ranging between 1 and 0.5 for $\gamma$ between 1 and 2. (The Navarro-Frenk-White (NFW~\cite{Navarro:1996gj}) halo falls in between $\gamma=1$ and $\gamma=2$.) This means that up to a model-dependent $\mathcal{O}(1)$ factor that depends on the details of $\kappa_0(\theta)$, one actually has $\kappa_{\rm c}(\theta_{\rm E})\sim \kappa_{\rm c}(\theta_{\rm E})/\kappa_0(\theta_{\rm E})$, arising implicitly because lensing prescribes that we evaluate $\kappa_{\rm c}(\theta)$ specifically near $\theta_{\rm E}$, determined in turn by $\kappa_0(\theta)$. Thus, up to a model-dependent $\mathcal{O}(1)$ factor, the detectable  $\kappa_{\rm c}$ depends on the fractional column density of ULDM compared with the dominant column density of the halo near the projected Einstein angle. 

{\it Imaging errors:} 
Because the ULDM-induced convergence profile $\kappa_{\rm c}(\theta)$ is not exactly constant, the MSD is only approximate and imaging errors are produced. The magnitude of these errors can be roughly estimated by assuming spherical symmetry, and evaluating the change induced on $\theta_{\rm E}$. The exact MSD would have the same value of $\theta_{\rm E}$, satisfying $\alpha_0(\theta_{\rm E})=\theta_{\rm E}$, regardless of the value of $\kappa_{\rm c}$. For finite core we parameterize the potentially-observable imaging distortion by the fractional change
\be\alpha(\theta_{\rm E})&:=&\theta_{\rm E}\left(1+\delta_{\rm E}\right).\ee
For this definition one finds~\cite{Blum:2021oxj}
\be\label{eq:dE}\delta_{\rm E}&=&2\int_0^1dy\,y\,\left(\kappa_{\rm c}(y\,\theta_{\rm E})-\kappa_{\rm c}(\theta_{\rm E})\right).\ee
Mock data analysis~\cite{Blum:2021oxj,Birrer:2020tax} suggests that current data for TDCOSMO systems may allow access to $\delta_{\rm E}\gtrsim3\%$ or so, although systematic modeling degeneracies may make this a difficult task (see, e.g.~\cite{Teodori:2022ltt,Teodori:2023nrz}). 

Our estimates for the $H_0$ signal $\delta_H$ (Eq.~(\ref{eq:H0_bias})) and the imaging error $\delta_{\rm E}$ (Eq.~(\ref{eq:dE})) are crude approximations because they are obtained in the spherical limit, while realistic systems exhibit deviation from sphericity, and also because they refer to the Einstein radius $\theta_{\rm E}$ while TDCOSMO systems have some imaging information also slightly off of $\theta_{\rm E}$. We explore this issue to some extent when we present our results, where, in addition to $\delta_{\rm E}$, we report the maximal variation of $\kappa_{\rm c}(\vec\theta)$ over all $|\vec\theta|\lesssim\theta_{\rm E}$. We also consider related effects, encoded in shear and flexion terms, in App.~\ref{s:shear}. A more precise treatment of the lensing inference pipeline, likely using mock analyses along the lines of Ref.~\cite{Blum:2021oxj,Birrer:2020tax}, would become important once more comprehensive cosmological simulations including ULDM, CDM, and stars become available.

\subsection{An aside: multiple-source systems}
The core-MSD effect could also, in principle, be detected via angular diameter distance ratios in multiple-source systems~\cite{Teodori:2022ltt}. For $ N $ planes (ignoring lens-lens coupling for simplicity), the lens equation associated to the $i$th source is modified as
\begin{equation}
\vec{\beta}_i = \vec{\theta} -  C_{i} \vec{\alpha}(\vec{\theta}) \ , \ C_i := \frac{D_{\rm S}  D_{\mathrm{LS}_i}}{ D_{\mathrm{S}i}  D_{\rm LS}}  \ ,
\end{equation}
where $ D_{\rm S,L,LS} $ are the angular diameter distances to source, lens, and lens-source, respectively, and $ z_{\rm S} $, $ z_{\rm L} $ are the redshifts. 
A core-MSD term in the main lens is observationally equivalent to an effective rescaling of the factors 
\begin{equation}
C'_{i} \approx C_{i} (1 + \delta\kappa^{\rm S}_{1i} -  \delta\kappa^{\rm LS}_{1i} + \kappa_{\rm c}(1-C_i)) \ ,
\end{equation}
where $ \delta\kappa^{\rm S,LS}_{1i} $ are differential external convergence terms, and the equality holds at first order in external convergence and $ \kappa_{\rm c} $. If one can constrain (e.g. from theory via cosmological simulations, or perhaps by combination with weak lensing data) the differential external convergence terms to be much smaller than the core factor $ \kappa_{\rm c}(1-C_i) $, and the redshifts and cosmological model parameters are measured well enough to determine the factors $C_i$ precisely (note that $H_0$ is not required for this purpose, because $H_0$ cancels out in $C_i$), then the core-MSD effect can be inferred from the difference between $C_i'$ and $C_i$. Preliminary analysis~\cite{Teodori:2022ltt} suggests that this method may be useful for $\kappa_{\rm c}\gtrsim0.1$ or so.

\section{Simulation setup} \label{s:setup}
We consider $10^{-25}\leq  m\leq10^{-24}$~eV, and simulate galaxy halos with halo mass $ M^{\rm halo}_{200}$ varying around $10^{13} M_\odot$, and cosmological ULDM fraction $ M_{\rm ULDM} \approx \alpha_\chi \, M^{\rm halo}_{200}$, fixed to $ \alpha_\chi=0.15 $ in most of the analysis. 

We use a pseudo-spectral Schr\"odinger-Poisson solver, based on the method of Ref.~\cite{Levkov:2018kau}, and implemented on a discretized grid with box length $ L\sim\SI{1}{\mega\parsec} $ varying slightly between different simulations. This box length is chosen as compromise between our computational resources and the need to resolve the wave dynamics of ULDM in $M^{\rm halo}_{200}\approx10^{13} M_\odot$ halos, which have $R_{200}$ of a few hundred kpc. More details on the code are deferred to App.~\ref{s:code}. 

Since ULDM makes-up only a subdominant fraction of the total DM, we need to implement also the dominant CDM component, as well as the stellar mass component in the halo (in TDCOSMO-like lens galaxies, the stellar mass is comparable to the DM mass near the projected Einstein radius). Our simulations cannot yet handle cosmological initial conditions, nor approximate realistic stellar physics. We therefore explore the sensitivity of the results to different approximate prescriptions for treating the galaxy halo: 
\begin{enumerate}[label=\Alph*.]
\item {\it CDM and stars as static external potential:} Our main results are obtained in a simple set-up, in which we approximate the dominant CDM and stellar mass distributions of the lens galaxy by fixed static external potentials. In this case, the numerical simulation includes only the ULDM dynamics, subject to the influence of the external potential. 

We approximate the CDM component, with mass $ M^{\rm CDM} = (1-\alpha_\chi)M^{\rm halo}_{200} $, as an NFW potential. The NFW scale radius parameter $r_{\rm s}$ in simulations is taken from the mass-concentration relations of~\cite{Maccio:2008pcd}.
The stellar component is taken to follow a Hernquist potential~\cite{Hernquist:1990}. 

\item {\it CDM as a heavier field with $m_{\rm heavy}\gg m$:} Here we approximately implement CDM by adding to the code another dynamical ULDM field, with a larger particle mass $m_{\rm heavy}\gg m$; for simplicity, we continue to refer to this component as ``CDM". The two fields interact with each other only via gravity.
The approximation of CDM by a large-$m$ Schr\"odinger field relies on the Vlasov-Poisson correspondence~\cite{Mocz:2018ium}. 
%Two field runs use a grid of $ 512^3 $ points, resulting in spatial resolution of $ L_{\rm box}/512 \lesssim \SI{2}{\kilo\parsec} $. 
We maintain sufficient numerical resolution to capture the de Broglie wavelength of the more massive field~\cite{Li:2018kyk,May:2021wwp}. The highest mass we can afford to simulate is $ m_{\rm heavy}=10^{-23}$~eV. Hence the ratio between $ m_{\rm heavy} $ and the lighter ``proper ULDM" $ m $ in our analysis varies between $10\leq m_{\rm heavy}/m\leq100$.  
In these simulations, the stellar component is neglected. Thus these simulations cannot be used directly to study gravitational lensing results. The goal of this analysis is to test the effect of backreaction of the subdominant ULDM field on the dominant CDM.

Simulating the background halo CDM with a dynamical field allows us to also test the impact of different initial conditions, notably the effect of galaxy mergers.  
\end{enumerate}
%

%{\it Initial conditions:} 
%
In all simulations, the ULDM density distribution is initialized approximately as an $\alpha_\chi$-scaled copy of the dominant CDM density profile. Initial conditions are defined via the Eddington procedure~\cite{Widrow:1993qq,Lancaster:2019mde}, adapted to wave equations:
\begin{equation}
	\psi_i(x) = (\Delta v)^{3/2}\sum_{\vec{v}} \sqrt{f(\mathcal{E}(x,v))} \e^{\iu m_i\vec{x}\cdot\vec{v} + \iu \varphi_{\vec{v}}} \ ,
\end{equation}
where $ \psi_i $ is the Schr\"odinger field associated to $ m_i $ (the index $i$ refers to the proper ULDM field $m$ and also to the $m_{\rm heavy}$ implementation of ``CDM", where relevant),  $ \varphi_{\vec{v}} $ is the random phase associated with the Fourier component of velocity $ \vec{v}=\vec k/m_i $, $ \Delta v $ is the velocity resolution step available in the simulation, and 
\begin{align}
	&f(\mathcal{E}) = \frac{2}{\sqrt{8}\pi^2} \int_0^{\sqrt{\mathcal{E}}} \dd{Q} \dv{^2\rho}{\Psi^2}{(Q)} \ , \\ 
	&\mathcal{E} = \Psi(r) - \frac{v^2}{2}  \ , \ \Psi = -\Phi + \Phi(r_{\rm max})  \ ,
\end{align}
where $\rho$ is the mass density of the simulated field, $ \Phi $ is the total gravitational potential and $  Q = \sqrt{\mathcal{E}-\Psi} $. The random phases $ \varphi_{\vec{v}} $ cause interference patterns that make runs with different random seeds differ from each other on scales smaller than $\lambda_{\rm dB}$, even if they are designed to give the same density profile on average.

Unless stated otherwise, simulations begin at $t=0$ and final results are read out at the Hubble time $t=t_H=1/H_0$. 

\section{Results}\label{s:results}
\begin{figure}
	\centering
	\includegraphics[width=0.475\textwidth]{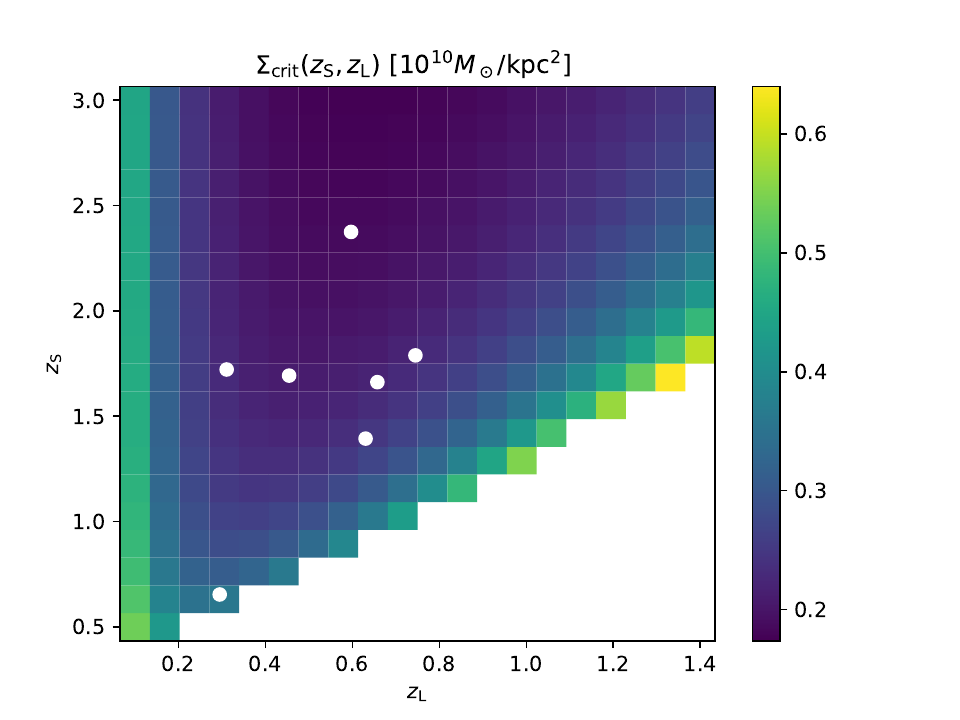}
	\caption{$ \Sigma_{\rm crit}(z_{\rm L}, z_{\rm S}) $ in the $(z_{\rm L},z_{\rm S})$ plane.  TDCOSMO systems~\cite{Millon:2019slk} are marked by dots.}
	\label{fig:sigma_crit}
\end{figure}

Our simulations produce results in terms of mass density profiles. However, gravitational lensing depends not only on the intrinsic density profile of the lens galaxy, but also on projection effects related to the observer-lens-source configuration.  In particular, mapping mass column density into lensing convergence requires the critical density $\Sigma_{\rm crit}=D_{\rm S}/\left(4\pi GD_{\rm LS} D_{\rm L}\right)$. 
Fig.~\ref{fig:sigma_crit} shows $ \Sigma_{\rm crit} $ in the $(z_{\rm L},z_{\rm S})$ plane (with the constraint $ z_{\rm S} > z_{\rm L} + 0.3  $). The seven lens systems from TDCOSMO~\cite{Millon:2019slk} are marked by white dots. Their $\Sigma_{\rm crit}$ ranges from 1.8 to 3.4, with a mean of 2.4$\times10^9~{\rm M_\odot/kpc^2}$. In presenting results that require convergence, we either use a reference value of $ \Sigma_{\rm crit} = \SI{2e9}{M_\odot\per\kilo\parsec\squared} $, or refer to specific examples from the TDCOSMO sample. 

\subsection{Static external CDM and stars}
We begin with simulations in which the dominant CDM and stellar components of the lens galaxy are approximated using static external NFW+Hernquist potentials. The dynamical ULDM has a cosmological fraction of $\alpha_\chi=0.15$. The CDM NFW profile has $M^{\rm halo}_{200}\approx10^{13} M_\odot$ with scale radius $r_{\rm s}\approx70$~kpc. 
The stellar Hernquist profile has a total mass of $\approx7.5\times10^{11}~{\rm M_\odot}$ and scale radius $\approx7$~kpc. At $r=10$~kpc the enclosed stellar mass is roughly twice the enclosed DM mass, compatible with estimates for typical TDCOSMO systems.

\begin{figure}[h]
	\centering
	\includegraphics[width=0.5\textwidth]{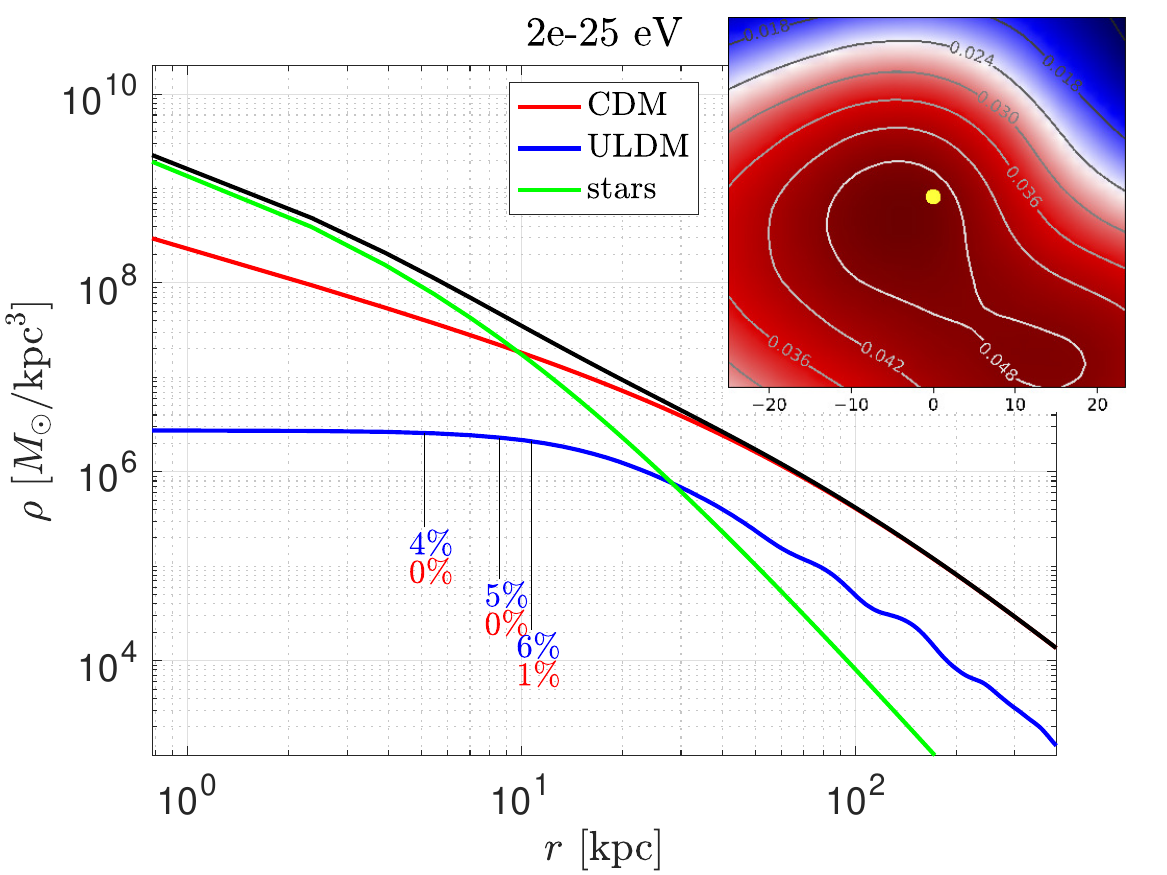}
	\caption{Radially-averaged density. Here $ m = \SI{2e-25}{\electronvolt} $, $\alpha_\chi=0.15$. Background halo implemented as static NFW+Hernquist potentials. We show $H_0$ signal ${\color{blue}\delta_H}$ (top number in {\color{blue}blue}) and imaging error ${\color{red}\delta_{\rm E}}$ (bottom number in {\color{red}red}) for three redshift configurations $(z_{\rm L},z_{\rm S})$, corresponding to real TDCOSMO systems. Each redshift configuration maps to slightly different projected Einstein radius $D_{\rm L}\,\theta_{\rm E}$. Colormap inset shows ULDM convergence $\kappa_{\rm c}$ using $\Sigma_{\rm crit}=2\times10^9~{\rm M_\odot/kpc^2}$ (grid units: kpc). Yellow spot marks the lens halo center of mass.
	}
	\label{fig:1}
\end{figure}

Fig.~\ref{fig:1} shows the radially-averaged 3D mass density for $ m = \SI{2e-25}{\electronvolt} $. 
We show $H_0$ signal ${\color{blue}\delta_H}$ (top number in {\color{blue}blue}) and imaging error ${\color{red}\delta_{\rm E}}$ (bottom number in {\color{red}red}) for three redshift configurations $(z_{\rm S},z_{\rm L})$, corresponding to TDCOSMO systems\footnote{From left to right in Fig.~\ref{fig:1}, these systems are RXJ1131, PG1115, and DESJ0408. Other TDCOSMO systems have $(z_{\rm L},z_{\rm S})$ combinations that place them in between these three in Fig.~\ref{fig:1}.}. Each redshift configuration maps to different projected Einstein radius $D_{\rm L}\,\theta_{\rm E}$, indicated on the plot by thin vertical lines connected to the corresponding ${\color{blue}\delta_H}$ and ${\color{red}\delta_{\rm E}}$. Colormap inset shows ULDM convergence $\kappa_{\rm c}$, using a typical reference value of $\Sigma_{\rm crit}=2\times10^9~{\rm M_\odot/kpc^2}$ to translate column density to lensing convergence (grid units: kpc). Yellow spot marks the halo center of mass.

Fig.~\ref{fig:1} showcases the AxionH0graphy scenario. ULDM is predicted to produce an MSD with $\kappa_{\rm c}(\theta_{\rm E})$ of a few percent, that would be essentially invisible in imaging and very challenging to detect via stellar kinematics alone (see discussion in Sec.~IV in Ref.~\cite{Blum:2021oxj}). Since the ULDM core falls off and joins the dominant CDM halo morphology after a few tens of kpc, it would also be difficult to detect in weak lensing, while cosmological constraints allow $\alpha_\chi=0.15$. Thus, if cosmography delivers $\lesssim2\%$ precision in inferred $H_0$~\cite{Birrer:2020jyr} when the MSD is {\it not} modeled explicitly, then Fig.~\ref{fig:1} demonstrates a physically-motivated scenario in which an ultralight axion may be first discovered from an apparent $H_0$ tension between the strong gravitational lensing measurement and other probes, e.g. early universe ones. %Perhaps this effect is already seen in data~\cite{Birrer:2020tax}.
%
%Fig.~\ref{fig:convergence5e25_extpot} shows a run with a $ m=\SI{5e-25}{\electronvolt} $ field, in an external NFW potential. The final snapshots of density profile and convergence are taken at $t=t_H$. Column density is converted to convergence using a fiducial value of the critical density $ \Sigma_{\rm crit} = \SI{0.2e10}{M_\odot\per\kilo\parsec\squared}  $, justified in Sec.~\ref{s:discussion}. 

\begin{figure*}
	\centering
        \includegraphics[width=0.425\textwidth]{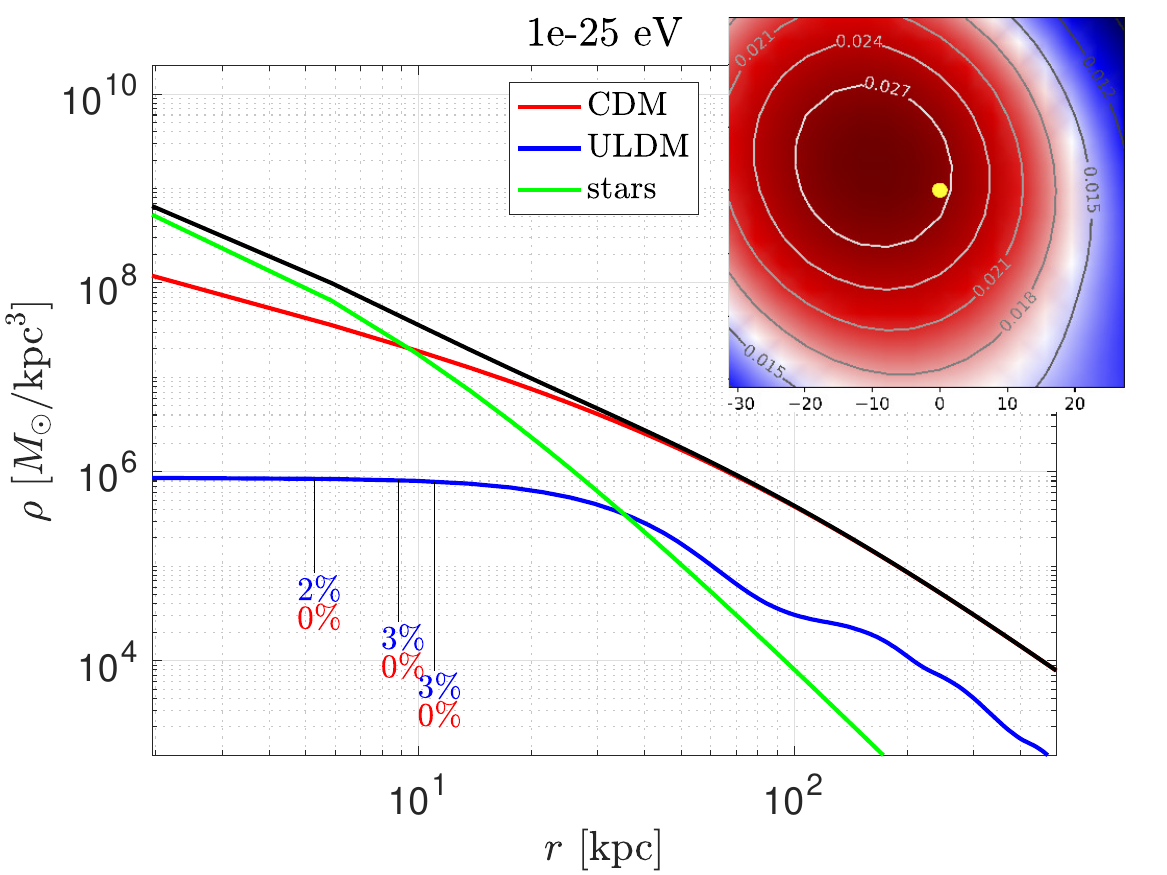}
	\includegraphics[width=0.425\textwidth]{plots/2e25-eps-converted-to.pdf}
	\includegraphics[width=0.425\textwidth]{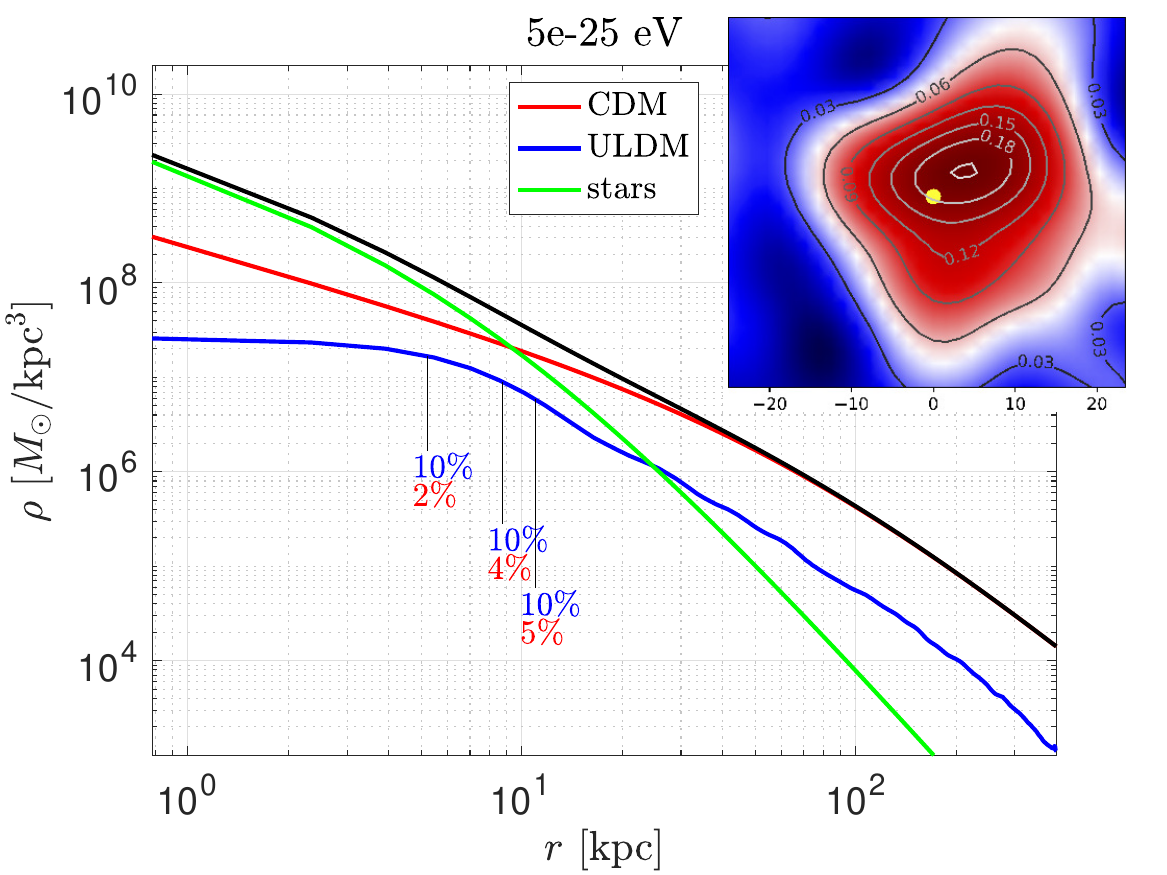}
	\includegraphics[width=0.425\textwidth]{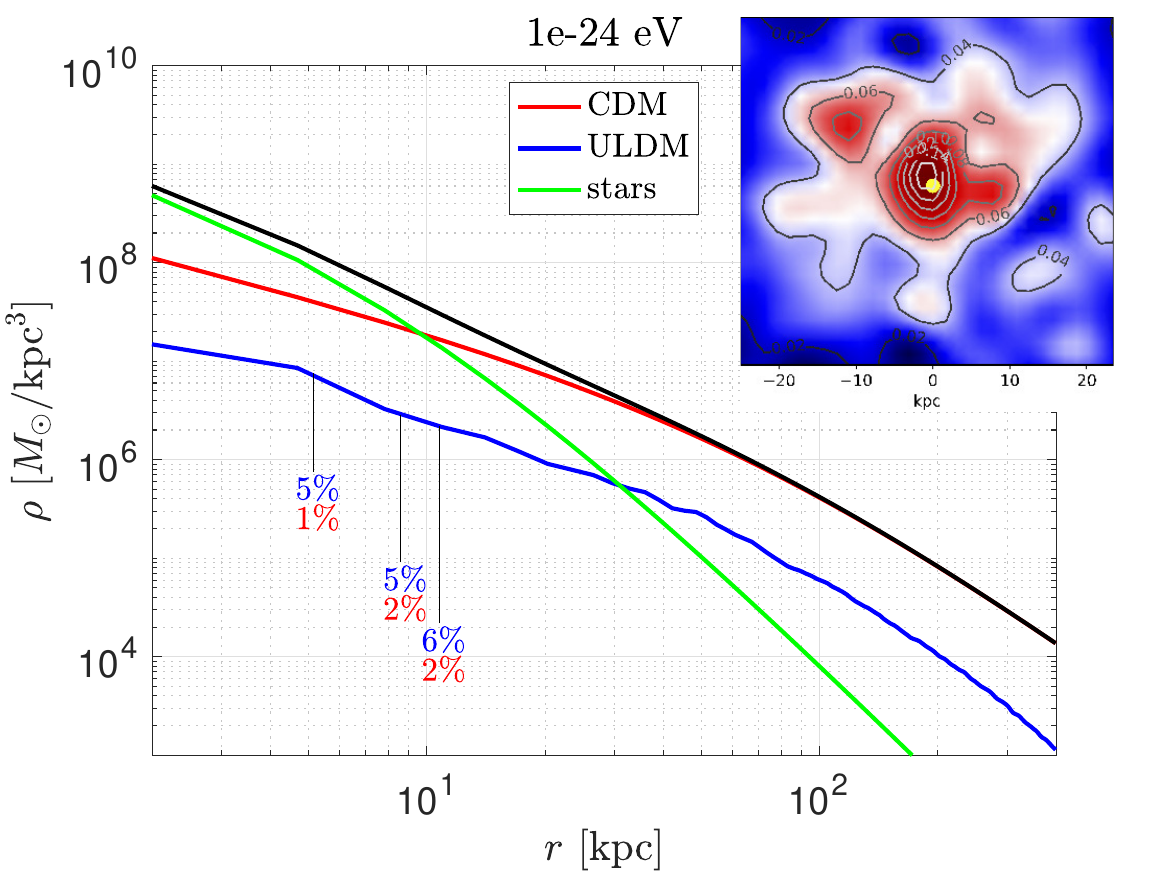}
	\caption{Same set-up as Fig.~\ref{fig:1} (reproduced in the top-right panel), extended to $ m = (1,5,10)\times10^{-25} $~eV (top-left, bottom-left, and bottom-right panels, respectively). 
	}
	\label{fig:12510e25}
\end{figure*}

Fig.~\ref{fig:12510e25} extends the results to $ m = (1,2,5,10)\times10^{-25}$~eV (this contains again the result of Fig.~\ref{fig:1} for comparison). %The value of $m$ for each panel is shown in the title. Ref.~\cite{Blum:2021oxj} anticipated that the range of $m$ where AxionH0graphy works would be roughly $10^{-25}\lesssim m\lesssim10^{-24}$~eV. As seen in Fig.~\ref{fig:12510e25}, and as we discuss below, this expectation is roughly consistent with our simulation results, albeit at the higher range of $m\gtrsim5\times10^{-25}$~eV wave interference and core displacement effects cause excess imaging distortions that require careful analysis. 

For $m=10^{-25}$~eV (top-left panel) the core extends far beyond the Einstein radius, making the MSD very precise. The ULDM-induced convergence is small -- about a factor of 5 smaller than $\alpha_\chi$ -- causing a small $H_0$ signal $\delta_H\lesssim3\%$ that is still, however, somewhat larger than cosmography precision expected in forecasts. It is interesting to also investigate the regime $m<10^{-25}$~eV, but our simulations do not extend there because of our technical limitation in simulating a sufficiently large box. 

$m\approx2\times10^{-25}$~eV (top-right panel, or Fig.~\ref{fig:1}) is a ``sweet-spot" for AxionH0graphy with $H_0$ signal $\delta_H\sim10\%$. 

For $m=5\times10^{-25}$~eV (bottom-left) the spatial scale of interference patterns in the ULDM field; the displacement of the ULDM core from the center of mass of the dominant CDM+stars halo~\cite{Li:2018kyk,Li:2020ryg}; and the core radius itself; which are all of the order of $\lambda_{\rm dB}$, approach the scale of the projected Einstein radius for a TDCOSMO-like galaxy. This causes deviations from the MSD approximation, seen in the bottom-left panel of Fig.~\ref{fig:12510e25} with $\delta_{\rm E}\gtrsim4\%$. In this regime the spherical approximation used in Eqs.~(\ref{eq:H0_bias}) and~(\ref{eq:dE}) becomes less accurate. 
Mock data analysis is needed to investigate this regime quantitatively. 
As a tentative rough estimate of the impact on imaging, noting that Eq.~(\ref{eq:dE}) is the spherical average of the difference between $\kappa_{\rm c}(\vec\theta)$ and its value at $\theta_{\rm E}$, we can compare $\delta_{\rm E}$ to the maximal variation in $\kappa_{\rm c}(\vec\theta)$ with respect to $ \kappa_{\rm c}(\theta_{\rm E}) $, taken across the region $|\vec\theta|\lesssim\theta_{\rm E}$: 
\begin{equation}\label{eq:Dkc}
\Delta\kappa_{\rm c} :=\max\limits_{|\vec\theta|<\theta_{\rm E}}(|\kappa_{\rm c}(\vec\theta) -  \kappa_{\rm c}(\theta_{\rm E})| )\ ,	
\end{equation}
where the coordinates are centered on the CDM halo center of mass.
Choosing $D_{\rm L}\,\theta_{\rm E}\approx10$~kpc for the boundary of the region, and using $\Sigma_{\rm crit}=2\times10^9{\rm M_\odot/kpc}$ for reference, we find $\Delta\kappa_{\rm c}\approx0.09$ for the $m=5\times10^{-25}$~eV simulation. Thus ULDM interference patterns and core displacement cause imaging errors of comparable magnitude to those caused by the finite-core effect. 
%(KB: relation to ~\cite{Xu:2013kna,Laroche:2022pjm,Powell:2023jns})

For $m=10^{-24}$~eV (bottom-right), the core radius is smaller than $D_{\rm L}\,\theta_{\rm E}$ and the convergence map shows that the MSD approximation is poor, violating Eqs.~(\ref{eq:H0_bias}) and~(\ref{eq:dE}) at the $\mathcal{O}(1)$ level. 
%We find a value of $\Delta\kappa_{\rm c}\approx0.05$, larger than the spherical estimate of $\delta_{\rm E}$, confirming that $\delta_{\rm E}$ ceases to be a reasonable indicator of imaging errors in this case.
%
%We emphasize that although AxionH0graphy is no longer a straightforward exercise for $m\gtrsim5\times10^{-25}$~eV, in the sense that the lensing time-delay $H_0$ bias may cease to be the key observable, this regime for $m$ is still very interesting for strong lensing, but it would require dedicated data analyses. 

\subsection{Dynamical CDM via $m_{\rm heavy}\gg m$}
\begin{figure*}
	\centering
	\includegraphics[width=0.32\textwidth]{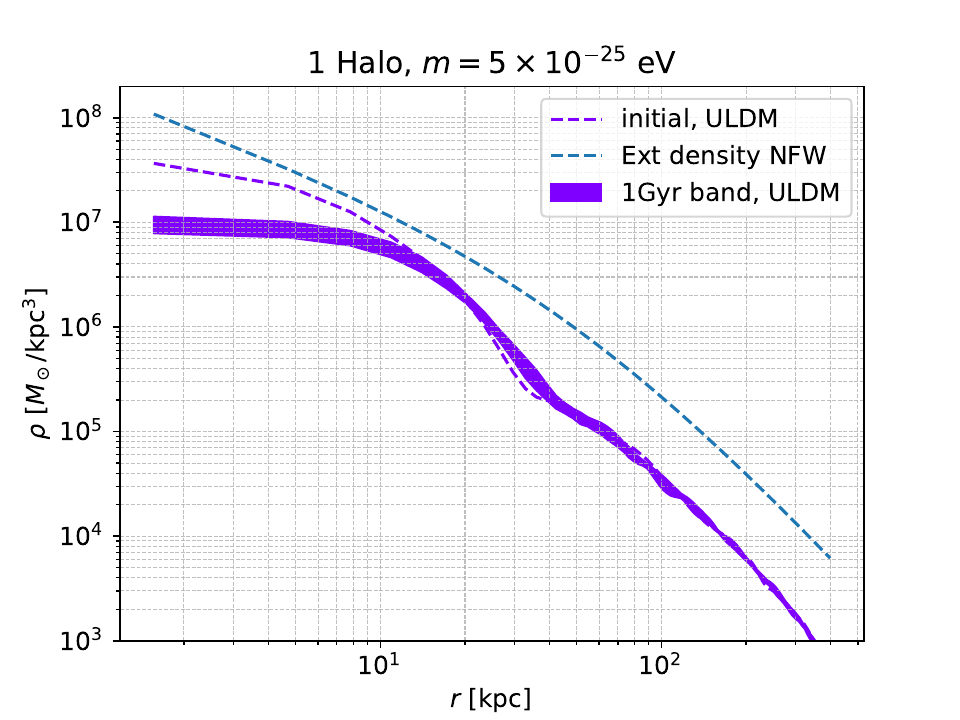}
	\includegraphics[width=0.32\textwidth]{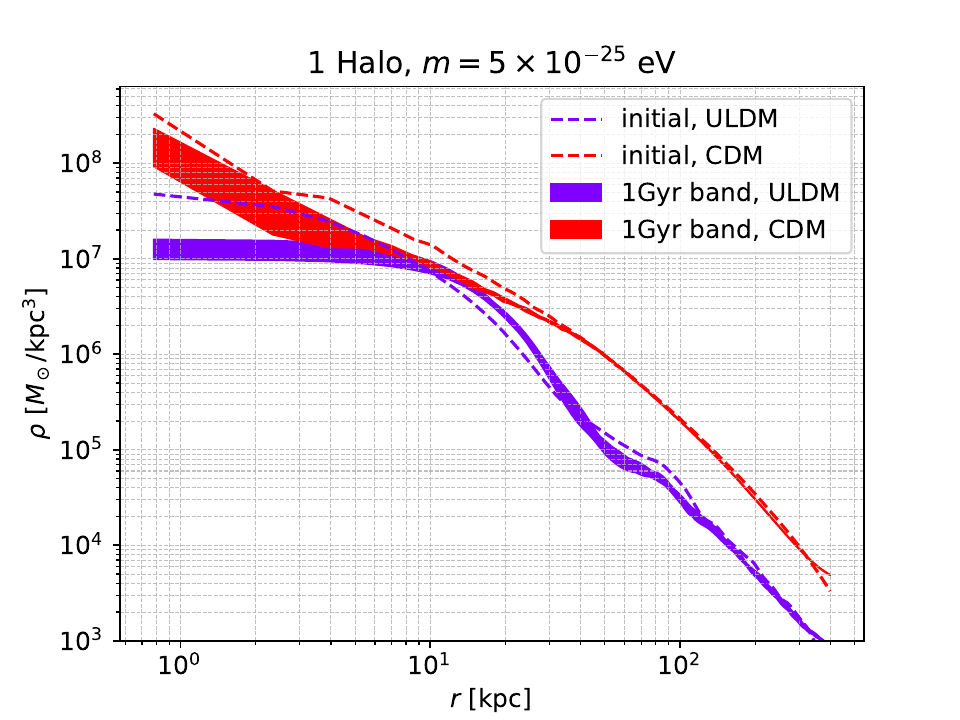}
	\includegraphics[width=0.32\textwidth]{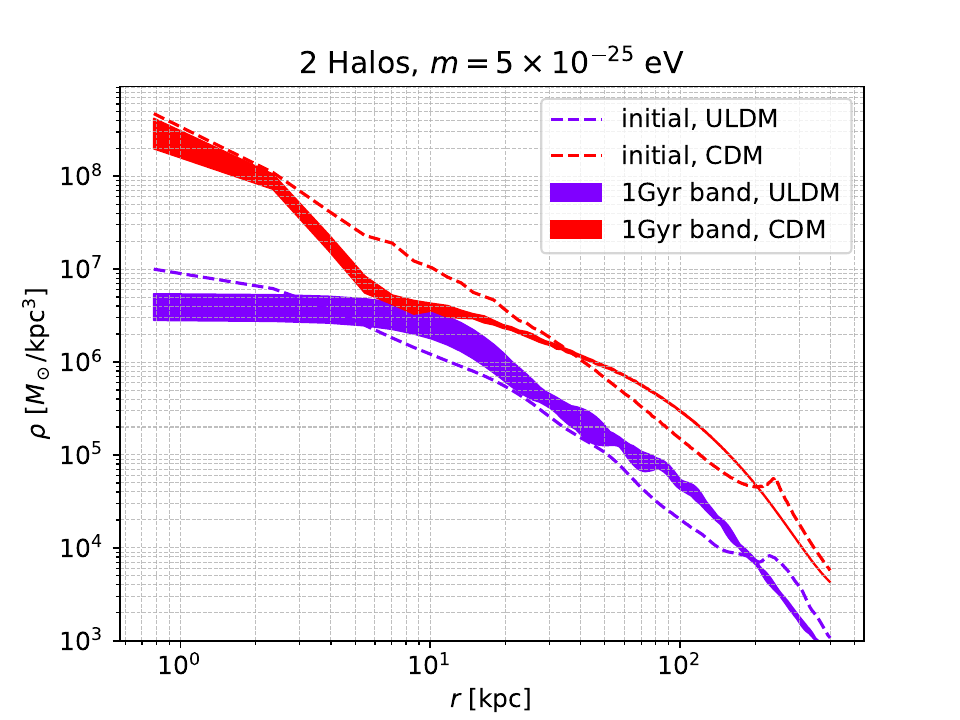}
	\caption{Simulation results for $ m = \SI{5e-25}{\electronvolt} $, $\alpha_\chi=0.15$. Panels differ in the implementation of the CDM halo component, and in initial conditions. Left panel: CDM implemented as static  NFW potential. Middle panel: CDM implemented by another dynamical field with $m_{\rm heavy}=10^{-23}$~eV. Right panel: CDM implemented as $m_{\rm heavy}=10^{-23}$~eV, initialized as two halos that merge dynamically. Final time results are shown by thick bands that contain the extreme values of the density field during the time $t=t_{\rm H}\pm1$~Gyr, capturing the effect of core oscillations. Initial conditions are shown by dashed lines. 
	}
	\label{fig:density5e25}
\end{figure*}

The middle panel of Fig.~\ref{fig:density5e25} shows simulation results for $m=5\times10^{-25}$~eV, where the CDM component of the halo is simulated by a dynamical field with $m_{\rm heavy}=10^{-23}$~eV. Initial conditions are shown by dashed lines. Final time results are shown by thick bands that contain the extreme values of the density field during the time $t=t_{\rm H}\pm1$~Gyr. The thickness of these bands demonstrates the amplitude of core oscillations and dynamical interference patterns. (Although we did not highlight this fact, comparable core oscillations exist in the simulations of Fig.~\ref{fig:12510e25}.)

For reference, the left panel of Fig.~\ref{fig:density5e25} shows results obtained with a fixed static NFW background, which approximately matches the CDM initial conditions in the dynamical case. These simulations do not include a stellar mass component: the goal  is to investigate the impact of the dynamical background on the ULDM core in a simple toy model. 

To test the sensitivity of our results to initial conditions, notably major mergers, we also perform runs initialized as two  NFW halos of similar initial mass profiles that merge dynamically. The two halos are initially separated by $ \approx \SI{100}{\kilo\parsec} $, at rest w.r.t. each other. The result is shown in the right panel of Fig.~\ref{fig:density5e25}.

Fig.~\ref{fig:density5e25} suggests that ULDM cores are not very sensitive to the details of CDM backreaction or halo initial conditions. In all three panels, the ULDM core central density is similar to within about a factor of two, with a similar core radius of about 20~kpc. 
Note that our simulations can only roughly resolve the detailed dynamics of the $m_{\rm heavy}=10^{-23}$~eV ``CDM" component, which exhibits intereference patters with typical wavelength of order 1~kpc. In particular, the right panel in Fig.~\ref{fig:density5e25} shows that the $m_{\rm heavy}$ field is forming a soliton of its own in the inner few kpc region of the halo. This could be interesting as it is expected that the lighter ULDM species, despite being subdominant in the total mass budget, could accelerate dynamical relaxation in the dominant CDM component~\cite{Bar:2021kti}. For AxionH0graphy, however, it is sufficient to notice that the ULDM core is not strongly dependent on the details of the main CDM component. 

%%%%%%%%%%%%%%%%%%%%
%
\begin{table*}
	\begin{tabular}{ccccccccccc}
\toprule
	$ \frac{m}{10^{-25}\text{eV}} $ &  $ \frac{M_{200}}{10^{12} M_\odot} $& $\frac{\rho_{\rm sol}(0)}{10^3\si{M_\odot\per\parsec\cubed}}$ &$ \frac{M_{\rm sol}}{10^{11} M_\odot} $
	& $ \frac{r_{\rm c}}{\text{kpc}} $ & $ \frac{\lambda_{\rm dB}}{2\text{kpc}} $ & $\frac{r_{\rm E}}{\text{kpc}}$ & $10^2\delta_{\rm E}$ 
	& $ 10^2\Delta \kappa_{\rm c} $ & $  \delta_H\approx\kappa_{\rm c}(\theta_{\rm E})  $  & Conditions \\
	\midrule
$1$ & $13.6$ & $0.4^{+0.1}_{-0.0}$ & $2.6^{+1.1}_{-0.0}$ & $37^{+8}_{-0}$ & $45$ & $8.4^{+3.1}_{-3.1}$ & $0.6^{+0.1}_{-0.2}$ & $0.3^{+0.1}_{-0.2}$ & $0.01^{+0.00}_{-0.01}$ & 1 halo, ext NFW \\
$1$ & $13.6$ & $0.9^{+0.3}_{-0.2}$ & $3.4^{+1.1}_{-0.0}$ & $33^{+4}_{-0}$ & $45$ & $8.4^{+3.1}_{-3.1}$ & $1.5^{+0.2}_{-0.6}$ & $0.4^{+0.3}_{-0.3}$ & $0.03^{+0.00}_{-0.01}$ & 1 halo, ext NFW + Hernq. \\
$1$ & $11.1$ & $0.4^{+0.2}_{-0.1}$ & $5.3^{+6.6}_{-0.0}$ & $48^{+12}_{-2}$ & $48$ & $4.4^{+2.2}_{-2.2}$ & $0.8^{+0.3}_{-0.2}$ & $0.1^{+0.1}_{-0.0}$ & $0.01^{+0.00}_{-0.00}$ & 1 halo, 2 fields \\
$1$ & $45.6$ & $6.2^{+4.8}_{-2.0}$ & $33.2^{+0.0}_{-17.0}$ & $27^{+9}_{-2}$ & $30$ & $26.7^{+9.0}_{-9.0}$ & $1.9^{+1.3}_{-0.2}$ & $\textcolor{red}{7.6^{+6.3}_{-8.0}}$ & $0.13^{+0.05}_{-0.08}$ & 1 halo, 2 fields \\
$1$ & $5.3$ & $0.2^{+0.0}_{-0.1}$ & $4.0^{+0.0}_{-1.6}$ & $56^{+23}_{-0}$ & $62$ & $4.8^{+1.8}_{-1.8}$ & $0.4^{+0.2}_{-0.1}$ & $0.0^{+0.0}_{-0.0}$ & $0.01^{+0.00}_{-0.00}$ & 2 halos, 2 fields \\
$1$ & $38.0$ & $4.8^{+7.5}_{-1.0}$ & $38.2^{+7.4}_{-6.7}$ & $38^{+2}_{-10}$ & $32$ & $18.3^{+4.7}_{-4.7}$ & $2.6^{+1.0}_{-0.6}$ & $\textcolor{red}{5.1^{+6.3}_{-4.5}}$ & $0.21^{+0.03}_{-0.10}$ & 3 halos, 2 fields \\
\midrule
$2$ & $21.0$ & $15.2^{+4.4}_{-6.7}$ & $7.7^{+3.2}_{-0.0}$ & $17^{+3}_{-0}$ & $20$ & $15.7^{+5.0}_{-5.0}$ & $2.0^{+0.4}_{-0.7}$ & $\textcolor{red}{5.5^{+2.6}_{-3.4}}$ & $0.12^{+0.02}_{-0.05}$ & 1 halo, ext NFW \\
$2$ & $12.8$ & $2.7^{+1.9}_{-0.0}$ & $5.9^{+0.0}_{-2.1}$ & $20^{+2}_{-0}$ & $23$ & $7.9^{+2.9}_{-2.9}$ & $0.8^{+0.5}_{-0.1}$ & $0.9^{+1.2}_{-0.7}$ & $0.05^{+0.01}_{-0.02}$ & 1 halo, ext NFW + Hernq. \\
$2$ & $7.8$ & $0.8^{+0.8}_{-0.0}$ & $6.5^{+0.0}_{-2.2}$ & $36^{+0}_{-7}$ & $27$ & $3.5^{+1.7}_{-1.7}$ & $3.0^{+2.5}_{-0.1}$ & $0.2^{+0.5}_{-0.2}$ & $0.05^{+0.00}_{-0.02}$ & 1 halo, 2 fields \\
$2$ & $24.0$ & $14.5^{+4.5}_{-2.3}$ & $14.7^{+0.0}_{-0.7}$ & $22^{+0}_{-2}$ & $19$ & $15.4^{+5.4}_{-5.4}$ & $\textcolor{red}{3.3^{+1.3}_{-0.8}}$ & $\textcolor{red}{12.0^{+10.1}_{-10.4}}$ & $0.24^{+0.07}_{-0.13}$ & 1 halo, 2 fields \\
$2$ & $21.0$ & $9.3^{+8.4}_{-5.5}$ & $10.7^{+10.3}_{-1.4}$ & $23^{+3}_{-5}$ & $20$ & $11.2^{+3.8}_{-3.8}$ & $2.6^{+1.2}_{-0.5}$ & $\textcolor{red}{7.0^{+8.9}_{-6.2}}$ & $0.20^{+0.04}_{-0.10}$ & 2 halos, 2 fields \\
$2$ & $11.3$ & $1.5^{+4.6}_{-0.0}$ & $14.6^{+0.0}_{-6.0}$ & $34^{+0}_{-9}$ & $24$ & $7.7^{+2.1}_{-2.1}$ & $1.3^{+0.9}_{-0.1}$ & $1.5^{+0.9}_{-1.0}$ & $0.07^{+0.01}_{-0.03}$ & 3 halos, 2 fields \\
\midrule
$5$ & $5.4$ & $7.9^{+3.4}_{-0.1}$ & $3.0^{+0.0}_{-0.8}$ & $14^{+3}_{-0}$ & $12$ & $4.2^{+1.6}_{-1.6}$ & $\textcolor{red}{10.9^{+2.9}_{-3.5}}$ & $1.9^{+3.7}_{-1.9}$ & $0.13^{+0.01}_{-0.06}$ & 1 halo, ext NFW \\
$5$ & $13.4$ & $25.9^{+19.1}_{-7.3}$ & $2.3^{+0.0}_{-0.6}$ & $7^{+3}_{-0}$ & $9$ & $8.4^{+3.0}_{-3.0}$ & $0.3^{+0.3}_{-0.0}$ & $\textcolor{red}{9.1^{+7.8}_{-7.8}}$ & $0.15^{+0.05}_{-0.08}$ & 1 halo, ext NFW + Hernq. \\
$5$ & $5.5$ & $15.9^{+0.0}_{-5.9}$ & $3.2^{+0.4}_{-0.0}$ & $12^{+3}_{-0}$ & $12$ & $4.7^{+1.8}_{-1.8}$ & $\textcolor{red}{6.4^{+6.3}_{-0.4}}$ & $\textcolor{red}{4.0^{+7.0}_{-3.5}}$ & $0.18^{+0.03}_{-0.08}$ & 1 halo, 2 fields \\
$5$ & $5.2$ & $1.1^{+2.3}_{-0.0}$ & $1.9^{+0.1}_{-1.0}$ & $18^{+0}_{-5}$ & $12$ & $4.5^{+1.8}_{-1.8}$ & $1.1^{+1.3}_{-0.2}$ & $0.7^{+1.1}_{-0.6}$ & $0.03^{+0.01}_{-0.02}$ & 1 halo, 2 fields \\
$5$ & $7.6$ & $3.3^{+2.1}_{-0.3}$ & $1.2^{+1.6}_{-0.0}$ & $15^{+2}_{-2}$ & $11$ & $6.1^{+2.1}_{-2.1}$ & $1.1^{+0.9}_{-0.0}$ & $2.6^{+2.3}_{-2.0}$ & $0.06^{+0.01}_{-0.03}$ & 2 halos, 2 fields \\
$5$ & $7.7$ & $7.4^{+0.8}_{-0.0}$ & $2.4^{+0.1}_{-0.1}$ & $15^{+0}_{-2}$ & $11$ & $6.3^{+1.6}_{-1.6}$ & $2.5^{+2.6}_{-0.2}$ & $2.1^{+3.2}_{-2.0}$ & $0.10^{+0.02}_{-0.05}$ & 3 halos, 2 fields \\
\midrule
$10$ & $9.9$ & $8.0^{+1.4}_{-0.0}$ & $0.4^{+0.0}_{-0.1}$ & $8^{+0}_{-0}$ & $5$ & $6.8^{+2.5}_{-2.5}$ & $1.9^{+1.8}_{-0.1}$ & $\textcolor{red}{7.1^{+4.7}_{-5.6}}$ & $0.10^{+0.04}_{-0.06}$ & 1 halo, ext NFW \\
$10$ & $12.8$ & $17.4^{+4.1}_{-1.5}$ & $0.2^{+0.0}_{-0.0}$ & $5^{+0}_{-0}$ & $5$ & $8.0^{+2.9}_{-2.9}$ & $2.4^{+1.3}_{-0.4}$ & $\textcolor{red}{3.1^{+2.4}_{-2.9}}$ & $0.06^{+0.02}_{-0.03}$ & 1 halo, ext NFW + Hernq. \\
$10$ & $7.6$ & $16.1^{+7.8}_{-6.0}$ & $1.0^{+1.1}_{-0.1}$ & $9^{+0}_{-2}$ & $5$ & $5.7^{+2.3}_{-2.3}$ & $0.2^{+0.1}_{-0.0}$ & $\textcolor{red}{4.5^{+3.5}_{-3.3}}$ & $0.10^{+0.02}_{-0.05}$ & 1 halo, 2 fields \\
$10$ & $8.4$ & $35.9^{+0.0}_{-11.4}$ & $2.1^{+0.8}_{-0.0}$ & $9^{+0}_{-0}$ & $5$ & $6.4^{+2.3}_{-2.3}$ & $\textcolor{red}{3.8^{+1.9}_{-0.7}}$ & $\textcolor{red}{11.0^{+12.8}_{-10.0}}$ & $0.22^{+0.07}_{-0.12}$ & 2 halos, 2 fields \\
$10$ & $12.8$ & $17.2^{+0.0}_{-6.5}$ & $0.8^{+1.3}_{-0.0}$ & $8^{+0}_{-0}$ & $5$ & $9.5^{+2.5}_{-2.5}$ & $0.9^{+0.3}_{-0.3}$ & $\textcolor{red}{4.4^{+1.4}_{-2.2}}$ & $0.08^{+0.01}_{-0.04}$ & 3 halos, 2 fields \\
	\bottomrule
\end{tabular}
\caption{Summary of simulation results for $\alpha_\chi=0.15$. From left to right: (1) ULDM particle mass, (2) $ M_{200} $ (final snapshot), (3) Central ULDM density (4) soliton mass from Eq.~\eqref{eq:fitsol}, (5) core radius, vis $  \rho_{\rm ULDM}(r_{\rm c}) = 0.5\rho_{\rm ULDM}(0)$, (6) half de-Broglie wavelength, using Eq.~\eqref{eq:deBroglie}, (7) projected Einstein radius $ r_{\rm E} =D_{\rm l}\theta_{\rm E}$, (8) $ \delta_{\rm E} $ from Eq.~\eqref{eq:dE} (in \%), (9) maximal convergence variation from Eq.~(\ref{eq:Dkc}) (in \%), (10) $H_0$ signal, or expected $\kappa_{\rm c}(\theta_{\rm E})$, (11) simulated configuration. }
\label{tab:runs}
\end{table*}

\subsection{Results summary for $\alpha_\chi=0.15$} \label{ss:discussion}
Tab.~\ref{tab:runs} summarizes the main results of our simulations for $\alpha_\chi=0.15$. For each of $m=(1,2,5,10)\times10^{-25}$~eV, we simulate a set of scenarios: single halo evolution with static CDM NFW potential; single halo with static CDM+stars; single halo with dynamical CDM implemented via $m_{\rm heavy}=10^{-23}$~eV~\footnote{For $m=(1,2,5)\times10^{-25}$~eV we report results of two separate runs.}; two-halo merger with dynamical $m_{\rm heavy}$ CDM; and three-halo merger with dynamical $m_{\rm heavy}$ CDM. 
All simulations   
are run until $t=t_H$. For each simulation we quote the CDM halo mass $M_{\rm 200}$ (extracted at $t=t_H$ when CDM is dynamical). We now discuss each additional column of Tab.~\ref{tab:runs} in turn.

{\it Soliton mass:} Ref.~\cite{Blum:2021oxj} parametrized the ULDM soliton profile in the presence of external potential with the form
\begin{equation}\label{eq:fitsol}
\rho_{\rm sol}(r) = \frac{\lambda^4}{(1+a^2\lambda^2 r^2)^{2b}}\frac{m^2}{4\pi G} \ ,% \ \lambda^4 := \rho_{\rm sol}(0) \ ,
\end{equation}
where $ a,b $ are fit parameters, and $\lambda$ a scaling variable. 
For simplicity, we use this parameterization even though the field configurations we find in the simulations deviate from spherical symmetry. 
Fitting the core with Eq.~\eqref{eq:fitsol}, we find the soliton mass up to $R_{200}$ as
\begin{equation}
M_{\rm sol} = 4\pi \int_0^{R_{200}} \dd{r} r^2 \rho_{\rm sol}(r) \ .
%\frac{\lambda}{Gm} \frac{\sqrt{\pi}\Gamma(-3/2 +2b )}{4a^3 \Gamma(2b)} \ .
\end{equation}
The error bars we quote for $ M_{\rm sol} $ and $\rho_{\rm sol}(0)$ come from maximum and minimum value in the period $t_{\rm H} \pm \SI{1}{\giga\year}$.

{\it Core scale radius $r_{\rm c}$, and $\lambda_{\rm dB}$:} We denote the scale length of the core in simulations by $r_{\rm c}$, defined as the radius at which the ULDM density decreases by a factor of 2 from its peak value. 
$r_{\rm c}$ is comparable with $\lambda_{\rm dB}/2$, estimated by
\begin{equation} \label{eq:deBroglie}
\lambda_{\rm dB} %= \frac{2\pi}{mv} 
\approx \SI{20}{\kilo\parsec} \qty(\frac{\SI{180}{\kilo\meter\per\second}}{v})\qty(\frac{\SI{5e-25}{\electronvolt}}{m}) \ ,
\end{equation}
where $ v $ is the 1D velocity dispersion %estimated via
\begin{equation}\label{e:v}
v \approx \sqrt{\frac{G M_{200}}{3R_{200}}} \approx \SI{180}{\kilo\meter\per\second}\qty(\frac{M_{200}}{10^{13} M_\odot})^{1/3} \ ,
\end{equation}
using 
%\begin{equation}
$R_{200} \approx \SI{450}{\kilo\parsec} \qty(\frac{M_{200}}{10^{13} M_\odot})^{1/3}$.\footnote{In estimating $\lambda_{\rm dB}$, which is only linearly sensitive to the value of $v$, and moreover only referred to in rough parametric estimates, we use the mean value of $v$ in the virial region of the halo. Later when we discuss dynamical relaxation, we would need to be a bit more precise and to estimate $v$ self-consistently in the region $r\sim\lambda_{\rm dB}$.}

Again, the error bars we quote for $ r_{\rm c} $ come from maximum and minimum value in the period $t_{\rm H} \pm \SI{1}{\giga\year}$.

{\it Einstein radius $r_{\rm E}$:} It is useful to compute the projected Einstein radius $r_{\rm E}=D_{\rm L}\theta_{\rm E}$ for comparison with $ r_{\rm c} $. We expect precise MSD when $ r_{\rm E} \ll r_{\rm c} $. In simulations that do not include a stellar mass component, we assign a value for $\theta_{\rm E}$ by adding to the convergence computation a contribution from a Hernquist profile, scaling the stellar mass linearly with $ M_{200} $. The error bars we quote for $r_{\rm E}$ are intended to represent projection effects. They relate to the smallest and largest $\theta_{\rm E}$ values obtained across the seven TDCOSMO systems (see Fig.~\ref{fig:sigma_crit}, and compare also to the vertical thin lines in Figs.~\ref{fig:1} and~\ref{fig:12510e25}). 

{\it Imaging error estimates $\delta_{\rm E}$ and $\Delta\kappa_{\rm c}$:} We quote both $\delta_{\rm E}$ (Eq.~\eqref{eq:dE}) and $\Delta\kappa_{\rm c}$ (Eq.~(\ref{eq:Dkc})) as estimates of imaging errors. 
Error bars refer to maximum and minimum values obtained from the redshift configurations of the TDCOSMO sample.

We highlight in {\color{red}red} $\delta_{\rm E}$ and $ \Delta\kappa_{\rm c} $ entries that exceed $ 0.03 $, as these halos may exhibit detectable MSD-breaking imaging errors. Note again that the error estimates do not take into account degeneracies involving external shear and flexion~\cite{Teodori:2022ltt,Teodori:2023nrz}. 
%Free modeling parameters associated with these degeneracies may absorb part of the imaging errors even for $ \delta_{\rm E}$ or $\Delta\kappa_{\rm c} >0.03 $. 
We comment on this point in App.~\ref{s:shear}. 

{\it $ H_0 $ signal:} We use Eq.~\eqref{eq:H0_bias} to estimate $H_0$ signal, or equivalently the expected $\kappa_{\rm c}(\theta_{\rm E})$, from the radial-averaged convergence profile.
The error bars we show come from the maximum and minimum values of $\kappa_{\rm c}(\theta_{\rm E})$ obtained by the redshift configurations of TDCOSMO systems. 

Finally, it is important to note that the ULDM mass in our simulated halos is matched to the cosmological fraction $ \alpha_\chi $. If cosmological evolution causes significant suppression of the ULDM fraction on distance scales of $ \sim $ Mpc, the actual ULDM mass in the inner region of galaxies may be lower than $ \alpha_\chi $. Our results could then exaggerate the detectability prospects of subdominant ULDM. The results in this work must be taken as approximations, which need to be validated via more refined simulations. The AxionH0graphy scenario may be misguided, if the approximations used to obtain our results prove to significantly overestimate the expected amount of $\kappa_{\rm c}$ in lens galaxies. This adds motivation to extend the analysis to full cosmological simulations which include $N$-body interactions for CDM and stars in future work.

\subsection{Brief view of $\alpha_\chi<0.15$}
Our work focused on $\alpha_\chi=0.15$, which is roughly the maximal ULDM fraction consistent with cosmological data~\cite{Kobayashi:2017jcf,Lague:2021frh}. Here we take a brief glance at smaller $\alpha_\chi$.

\begin{figure}
	\centering
	\includegraphics[width=0.45\textwidth]{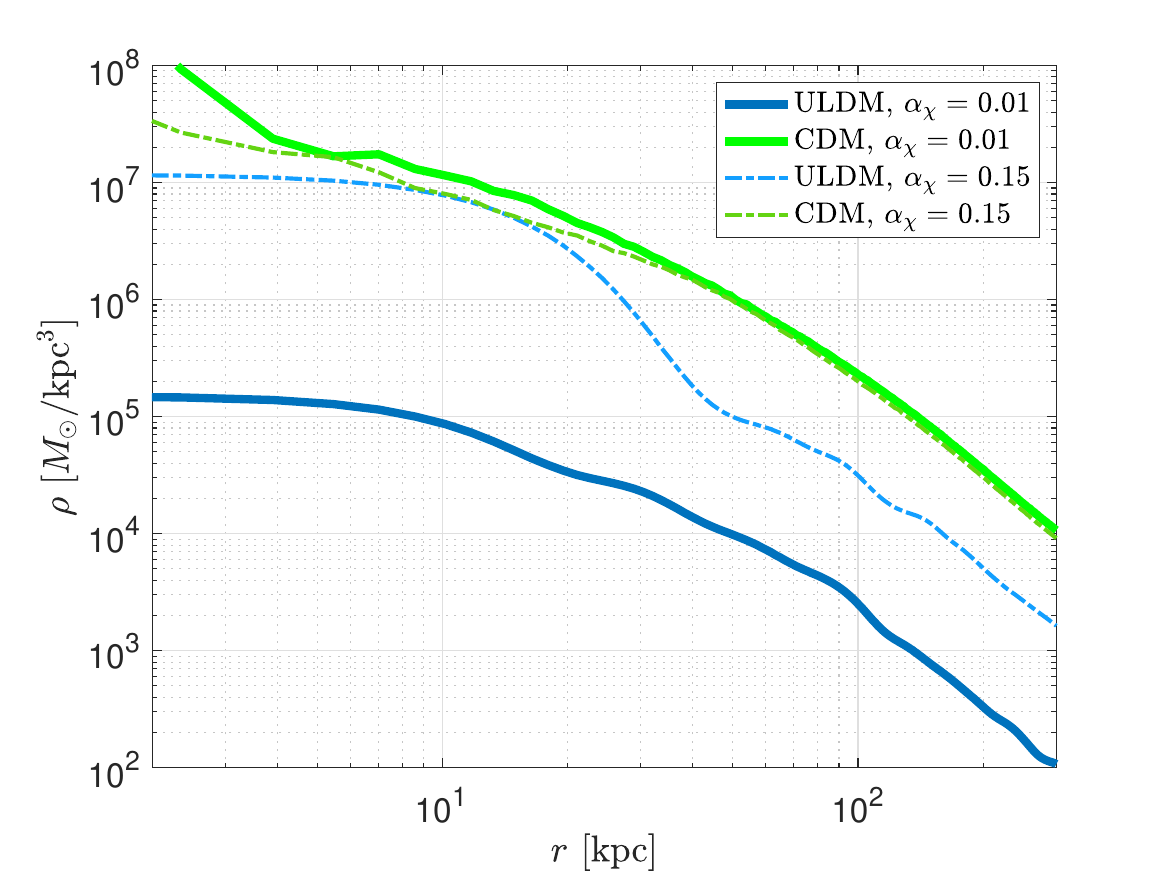}
	\caption{Comparison of results for $\alpha_\chi=0.01$ (thick solid lines) and $\alpha_\chi=0.15$ (thin dot-dashed lines), done for $m=5\times10^{-25}$~eV, with dynamical CDM modeled via $m_{\rm heavy}$ field. 
	}
	\label{fig:alpha5e25}
\end{figure}

Fig.~\ref{fig:alpha5e25} compares simulations for $\alpha_\chi=0.01$ (thick solid lines) and $\alpha_\chi=0.15$ (thin dot-dashed lines), done for $m=5\times10^{-25}$~eV. Results for $\alpha_\chi=0.15$ are the same as in the middle panel of Fig.~\ref{fig:density5e25}. The comparison highlights the soliton core that begins to form in the $\alpha_\chi=0.15$ case, but not in the $\alpha_\chi=0.01$ case, in qualitative agreement with  results in~\cite{Lague:2023wes,Schwabe:2020eac}. 

\subsection{Comments on stellar kinematics}
Stellar kinematics measurements can be used to constrain the MSD. 
As of now, works like Ref.~\cite{Birrer:2020tax,Shajib:2023uig} attempt to measure $H_0$ by mitigating the MSD effects via stellar kinematics. As we already mentioned, a mild positive evidence (albeit not statistically significant) for the presence of mass sheets arises on Ref.~\cite{Birrer:2020tax}. In the AxionH0graphy scenario, precise stellar kinematics, alongside time delays, are both used to detect internal MSD effects, potentially arising from large ULDM cores. 

Currently, stellar kinematics measurements allow for an order $10 \%$ exact MSD. Moreover, an approximate MSD with finite core radius has a {\it smaller} impact on stellar kinematics, when compared to the exact MSD. We showed this point, using spherical Jeans analysis, in Ref.~\cite{Blum:2021oxj}, Sec.~IV there. This makes time delays a complementary and currently more precise method to stellar kinematics for probing large cores.

\section{Timescale of core formation}\label{s:t}
So far we focused on simulation snapshots at $t=t_H$. We now comment on the time evolution leading to these snapshots. 

Fig.~\ref{fig:growth} shows the time evolution of the peak ULDM density of two simulations, containing $m=2\times10^{-25}$~eV (left panel) and $5\times10^{-25}$~eV (right panel) at $\alpha_\chi=0.15$ with dynamical $m_{\rm heavy}$ CDM. We can identify three regimes in the dynamics of the core:
\begin{figure*}
	\centering
	\includegraphics[width=0.45\textwidth]{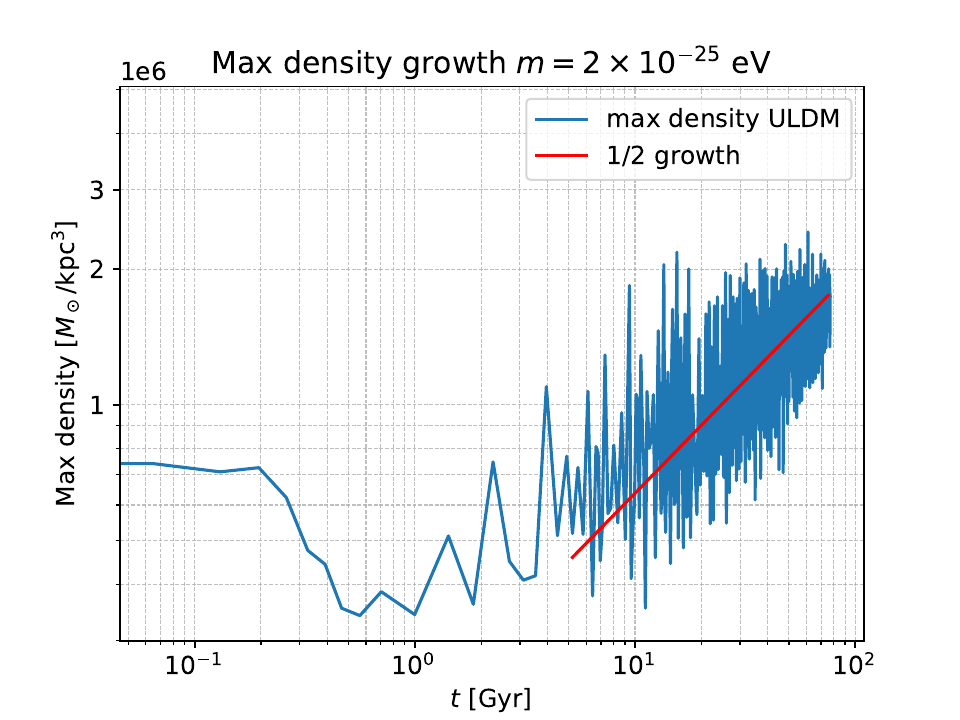}
 	\includegraphics[width=0.45\textwidth]{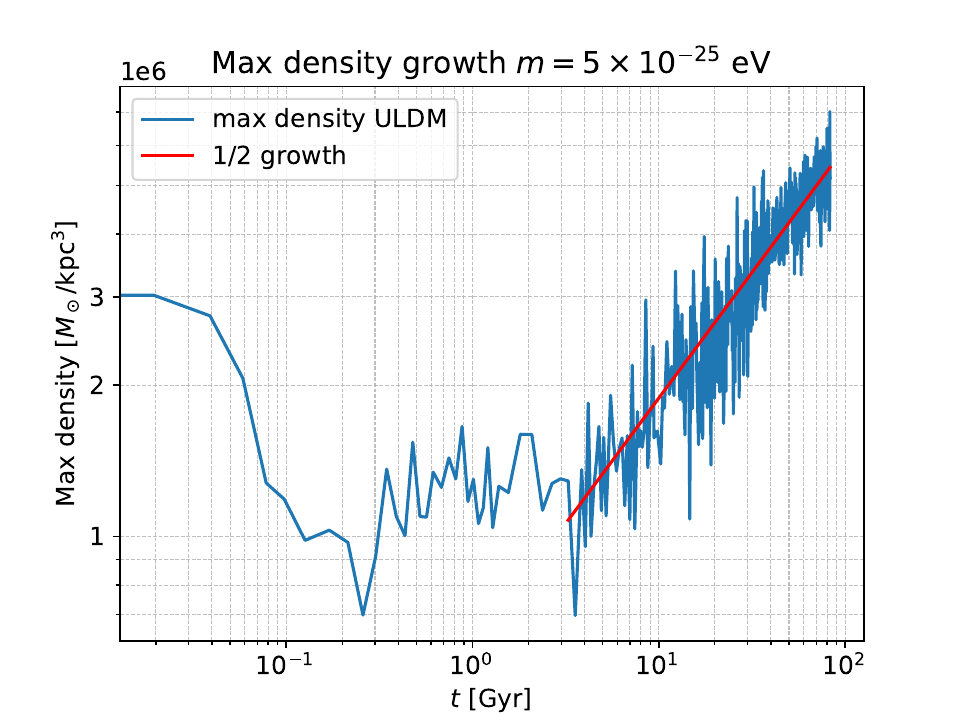}
	\caption{ULDM core density as a function of time. Left (right) panel: $m=2~(5)\times10^{-25}$~eV.
	}
	\label{fig:growth}
\end{figure*}

\begin{enumerate}
\item {\it Wave pressure expels mass outwards, forming flattened region.} As explained in Sec.~\ref{s:setup}, in the (simplistic) initial conditions of our simulations the ULDM halo is an approximate copy of the dominant CDM NFW halo, scaled-down by a factor of $\alpha_\chi$. 
%(if more than one halo is present in the initial condition, as in the case of two and three halo merger simulations, then each initial CDM halo comes along with its own ULDM component).  
%
The Eddington prescription used to initialize the field does not take into account wave mechanics pressure. This  omission is analogous  to neglecting two-body stellar encounters when assigning initial conditions in standard N-body simulations. 
Once we start the run, ULDM in the dense central region of the halo evolves quickly due to wave pressure to form a flattened core, on a wave-mechanics time scale $t_{\rm dB}$ given by 
\begin{align}
	\begin{aligned}
	t_{\rm dB}&=\frac{1}{mv^2}\\
	&\approx0.1\left(\frac{200~{\rm km/s}}{v}\right)^2\left(\frac{5\times10^{-25}{~\rm eV}}{m}\right)~{\rm Gyr} \ .	
	\end{aligned}
\end{align}
$t_{\rm dB}$ is comparable but probably somewhat shorter than the ($m$-independent) free-fall time,
\be
t_{\rm free fall} &=& \sqrt{\frac{3\pi R^3_{200}}{32 G M_{200}}} \approx \SI{0.8}{\giga \mathrm{yr}} \ .
\ee

We do not investigate the details of the initial short-time dynamics, because these necessarily depend on the details of our rather ad-hoc initial conditions. The only comment we make in this regard is that the later evolution of the core is relatively independent on the details of the initial conditions, as those are quickly ``ironed out" by the wave pressure. 

\item {\it Dynamical relaxation and emergence of a soliton.} 
After the initial density drop, the core density fluctuates for a while without substantial mean growth. 
%The core peak density is typically displaced from the halo center of mass by a distance of order $\lambda_{\rm dB}$, and the core exhibits random motion and oscillations around a relatively steady mean. 

The state of the field during this stage may not be very different from the random noise state used in Ref.~\cite{Levkov:2018kau,Bar-Or:2018pxz,Bar-Or:2020tys} to investigate soliton condensation in a statistically homogeneous background. The relaxation timescale in that case was found to be
\be\label{eq:treleq}
t_{\rm rel} &\approx& \frac{\sqrt{2}}{12\pi^3} \frac{m^3 v^6}{G^2\rho_\chi^2 \Lambda} \ ,
\ea
where $\rho_\chi$ refers to the ULDM density and $\Lambda$ is a Coulomb logarithm, of order unity near the core. The simulations of soliton condensation from random noise, done in Ref.~\cite{Levkov:2018kau}, show self-gravitating solitons emerging from noise around $t\approx t_{\rm rel}$, after which time the soliton mass grows approximately as $M_{\rm sol}\propto t^{\frac{1}{2}}$. 

To compare this expectation to our simulations, we can read $\rho_\chi$ directly from Fig.~\ref{fig:growth}. However, $t_{\rm rel}$ depends very sensitively on the local velocity dispersion $v$ in Eq.~(\ref{eq:treleq}), which is not a straightforward quantity to define precisely in our situation, where dispersion sourced by the background potential adds to the motion of the ULDM core around the halo center of mass. We make a crude estimate as follows. From the simulation, we read the core radius $r_{\rm c}$ of the ULDM; we find $r_{\rm c}\approx26~(18)$~kpc for $m=2~(5)\times10^{-25}$~eV, with enclosed background halo mass $M_{\rm halo}(r_{\rm c})\approx0.35~(0.2)\times10^{12}$~M$_\odot$. We then estimate the circular velocity at $r_{\rm c}$ as $v^2_{\rm circ}(r_{\rm c})=GM_{\rm halo}(r_{\rm c})/r_{\rm c}$, and use $v\approx\sqrt{3}\times v_{\rm circ}$ in Eq.~(\ref{eq:treleq}). This crude estimate gives $v_{\rm circ}(r_{\rm c})\approx245~(222)$~km/s, and $t_{\rm rel}\approx0.9~(8)$~Gyr, for $m=2~(5)\times10^{-25}$~eV, corresponding to the left (right) panel of Fig.~\ref{fig:growth}. This estimate is not very far from the behavior seen in Fig.~\ref{fig:growth}. 
However, to appreciate how sensitive these estimates are to numerical details, note that using $r_{\rm c}\to2\times r_{\rm c}$ in our estimate would multiply the resulting $t_{\rm rel}$ by a factor of $\sim8$. We can only conclude that our results in Fig.~\ref{fig:growth} are not clearly inconsistent with Eq.~(\ref{eq:treleq}).

\item {\it Soliton growth.} 
In the third stage of the evolution the central density in our simulations shows a clear trend of growth, consistent with $\rho_{\rm c}\propto t^{\frac{1}{2}}$. To guide the eye, we superimpose a red line with this slope in both panels. 

Note that we see $\rho_{\rm c}\propto t^{\frac{1}{2}}$, while Ref.~\cite{Levkov:2018kau} reported that the {\it mass} of the soliton, $M_{\rm sol}$, grows as $\propto t^{\frac{1}{2}}$. This scaling is consistent if the soliton is not self-gravitating, but rather dominated by the velocity dispersion due to the background potential, in which case the soliton radius only weakly depends on the soliton mass and one finds $M_{\rm sol}\propto\rho_{\rm c}$. This is indeed the relevant regime in our simulations.

We track the build-up of the soliton to $t\gg t_{\rm H}$. We do not see a clear break of the soliton growth, although we cannot exclude the beginning of a break at late time. Saturation of soliton growth is expected to occur when the velocity dispersion due to the soliton's self-gravity becomes comparable to the mean velocity dispersion in the halo (see discussion and references in~\cite{Bar:2018acw,Bar:2021kti}). The solitons in our simulations do not reach this point at $t=t_{\rm H}$; even the relatively prominent ULDM cores seen in the middle and right panels of Fig.~\ref{fig:density5e25}, where the ULDM density is comparable to the CDM density near the core radius, still induce a contribution to $v_{\rm circ}$ that is smaller by a factor of few compared to the CDM-dominated peak halo $v_{\rm circ}$. This result is consistent with expectations in Ref.~\cite{Blum:2021oxj} that identified soliton growth as a bottleneck for AxionH0graphy.

\end{enumerate}
%

%%%%%%%%%%%%%%%%%%%
\section{Summary}\label{s:sum}
Even a subdominant contribution of ultralight dark matter (ULDM), of order ten percent of the total DM budget, can affect strong gravitational lensing at a level accessible to current observations and makes a promising target for near-future surveys. This happens because ULDM condenses into cores around massive galaxies. We reported results of a series of numerical simulations investigating this effect, extending the original analysis of Ref.~\cite{Blum:2021oxj}.

For ULDM particle mass $m\sim10^{-25}$~eV, the scale size of the cores is a few tens of kpc in massive elliptical galaxies of the type used for  cosmographic determination of $H_0$. The main effect of a subdominant component of ULDM would be to cause a subtle positive bias in the inferred value of $H_0$. Adding an external prior on $H_0$ from another data set (e.g. CMB, or the supernovae distance ladder), the subdominant ULDM component can be detected. We dub this method AxionH0graphy. It provides concrete theoretical motivation for combining cosmography with an external $H_0$ prior to investigate the structure of massive galaxies, demonstrating the potential for fundamental discovery.%~\cite{Blum:2020mgu}.

$m\gtrsim5\times10^{-25}$~eV ULDM would introduce imaging distortions in addition, if not instead of the MSD time-delay shift. This regime for $m$ is also interesting but the emphasis shifts to detailed lens modeling, rather than an $H_0$ prior.

We believe our simulations provide robust results regarding the formation, general scale, and dynamics of ULDM cores. However, the set-up we explore is still simplistic in many important aspects: first, we cannot yet simulate fully self-consistently the dominant component of cold DM (CDM) and stars, that are both important for strong lensing in  massive elliptical galaxies. Our simulations either use static external potentials to replace these dominant backgrounds, or a wave implementation of CDM that cannot resolve the dynamics below $\sim1$~kpc scales. Second, our simulations start from rather ad-hoc initial conditions, and may miss some features of cosmological structure formation. 
More refined studies and simulations may either confirm or refute the theoretical expected effects of subdominant ULDM component on strong gravitational lensing, as presented in this work.

With these caveats in mind, our predictions for the amplitude of the subdominant ULDM effect on strong lensing systems must be understood as crude approximations, motivating more comprehensive analysis.

%%%%%%%%%%%%%%%%%%%
\acknowledgments
We are grateful to E. Hardy and M. Gorghetto for useful discussion and guidance in the use of the pseudo-spectral solver. Different versions of the code used for this work were also used in~\cite{Gorghetto:2022sue,Budker:2023sex,Gorghetto:2024vnp}.
We thank M. Nori and M. Baldi for collaboration in the early stage of this work. KB and LT were supported by ISF grant 1784/20 and by MINERVA grant 714123

\bibliography{ref}
\bibliographystyle{utphys}

\appendix
%\section{Notes on NFW}\label{a:nfwkappa}
%%%%%%%%%%
%For $\theta\ll\theta_{\rm s}$, the convergence due to an NFW halo with concentration parameter $c=R_{\rm vir}/r_{\rm s}$ 
%is given by
%%
%\be\kappa_{\rm nfw}(\theta)&\approx&\frac{2 r_{\rm s}\rho_{\rm c}\delta_{\rm c}}{\Sigma_{\rm crit}}\ln\left(\frac{2\theta_{\rm s}}{\e\theta}\right),\ee
%% 
%valid up to corrections of order $\mathcal{O}\left((\theta/\theta_{\rm s})^2\ln(\theta_{\rm s}/\theta)\right)$, 
%where we defined, as usual, $\delta_{\rm c}=\frac{200}{3}\frac{c^3}{\ln\left(1+c\right)-\frac{c}{1+c}}$. 
%%
%The deflection angle (again for $\theta\ll\theta_{\rm s}$) is 
%%
%\be\alpha(\theta)&=&\frac{2}{\theta}\int_0^\theta \dd{y}y\kappa_{\rm nfw}(y)\approx\frac{r_{\rm s}\rho_{\rm c}\delta_{\rm c}\theta}{\Sigma_{\rm crit}}\ln\left(\frac{4\theta_{\rm s}^2}{\e\theta^2}\right).\ee
%%
%With this, the Einstein angle is 
%%
%\be\theta_{\rm E}&\approx&2\theta_{\rm s}\e^{-\frac{1}{2}\left(\frac{\Sigma_{\rm crit}}{r_{\rm s}\rho_{\rm c}\delta_{\rm c}}+1\right)}.\ee
%%
%The convergence at the Einstein angle is
%%
%\be\kappa_{\rm nfw}(\theta_{\rm E})&=&1-\frac{r_{\rm s}\rho_{\rm c}\delta_{\rm c}}{\Sigma_{\rm crit}}.\ee
%%
%Above we assumed $\theta\ll\theta_{\rm s}$; indeed, actual TDCOSMO systems~\cite{Millon:2019slk} happen to have $\theta_{\rm E}\sim0.1\theta_{\rm s}$. This translates into $\frac{r_{\rm s}\rho_{\rm c}\delta_{\rm c}}{\Sigma_{\rm crit}}\sim0.2$ and $\kappa_{\rm nfw}(\theta_{\rm E})$ close to and slightly smaller than 1.
%
%The choice of $r_{\rm s}$ in simulations is based on mass-concentration relations of~\cite{Maccio:2008pcd}.

\section{Schr\"odinger-Poisson solver} \label{s:code}
%%%%%%%%
We use a pseudo-spectral 3D Schr\"odinger-Poisson solver, which closely resembles the one described in~\cite{Levkov:2018kau}. In the following, we review the basics of its implementations, together with some details of the particular current setup.

The Schr\"{o}dinger-Poisson Equations for $ n $ fields $ \psi_i $ with mass $ m_i $ read (in our conventions, $ \psi $ has the dimensions of a mass squared, in particular the density is $ \rho = |\psi|^2 $) %case with masses $ m_i $; denoting $ r_i = m_i/m_1 $, we have
\begin{align} \label{eq:SPEnfields}
	&\iu\pdv{\psi_i}{t} = -\frac{1}{2m_i} \laplacian{\psi_i} + m_i(\Phi+\Phi_{\rm ext})\psi_i \ , \\
& \laplacian \Phi = 4\pi G\sum_i(|\psi_i|^2 - \langle|\tilde\psi_i|^2|\rangle) \ ,		
\end{align}
where $ \Phi_{\rm ext} $ is a possible source of external gravitational potential, not sourced by $ \psi_i $, for which back-reaction is neglected. 

To convert to adimensional variables, define
\begin{equation} \label{eq:rescaling}
	\tilde{\psi}_i = \frac{1}{\beta} \frac{\sqrt{4\pi G}}{m_1} \psi \ , \ \tilde{x} = \sqrt{\beta} m_1 x \ , \ \tilde{t} = \beta m_1 t \ , \
	\tilde{\Phi} = \Phi/\beta \ ,
\end{equation}
with $\beta $ an adimensional rescaling parameter. 	
This gives
\begin{align}
	&\iu\pdv{\tilde{\psi_i}}{\tilde{t}} = -\frac{1}{2r_i} \laplacian_{\tilde{x}}{\tilde{\psi_i}} + r_i(\tilde{\Phi}+ \tilde{\Phi}_{\rm ext})\tilde{\psi_i} \ , \\
	& \laplacian_{\tilde{x}} \tilde{\Phi} = \sum_i|\tilde\psi_i|^2   - \langle\sum_i|\tilde\psi_i|^2|\rangle \ ,	
\end{align}
where we defined
\begin{equation}
	r_i := \frac{m_i}{m_1} \ .
\end{equation}
The energy associated to each field component reads 
%(be careful about our rescaling of the field $ \psi_i $, our Poisson equation is different with respect to Levkov one for example)
\begin{equation}
	\tilde{E}_i = \int \dd[3]{\tilde{x}} \qty( \frac{|\grad{\tilde{\psi_i}}|^2}{2r^2_i} + \frac{1}{2} |\tilde{\psi_i}|^2 \tilde{\Phi} +|\tilde{\psi_i}|^2 \tilde{\Phi}_{\rm ext} ) \ .
\end{equation}

The field is evolved in time via the unitary operator
\begin{equation}
\tilde\psi_i(\tilde{t}+ \dd \tilde{t}) = \prod_{\alpha} \e^{-\iu r_i d_\alpha \dd{\tilde{t}} (\tilde{\Phi}_\alpha + \tilde{\Phi}_{\rm ext} )} \e^{-\iu c_\alpha\dd{\tilde{t}} \frac{(-\iu \grad)^2}{2r_i} }\tilde\psi_i(\tilde{t}) \ .
\end{equation}
The equation is meant to be read from right to left, i.e., one first applies the kinetic operator
\begin{equation}\label{eq:kin_op}
\e^{-\iu c_\alpha\dd{\tilde{t}} \frac{(-\iu \grad)^2}{2r_i} }\tilde\psi_i(\tilde{t}) =: \tilde\psi^{(\alpha)}_i(\tilde{t} +\dd{\tilde{t}} ) \ ,
\end{equation}
and then the potential operator, where
\begin{equation}\label{eq:Poiss_num}
\laplacian{\tilde\Phi_\alpha} = \sum_i|\tilde\psi_i^{(\alpha)}|^2 \ .
\end{equation}
The constants $ c_\alpha $, $ d_\alpha $ can be found in~\cite{Levkov:2018kau}. Eqs.~\eqref{eq:kin_op},\eqref{eq:Poiss_num} are solved via Fast Fourier Transforms (FFT), using the FFTW library~\cite{10.1145/301631.301661}.

We use adaptive time-steps, ensuring conservation of energy $ \Delta E/E \lesssim 10^{-5} $ between time-steps, and overall conservation of energy $ |E^{\rm tot}_{\rm final} - E^{\rm tot}_{\rm initial}|/|E^{\rm tot}_{\rm final} + E^{\rm tot}_{\rm initial}| \lesssim 10^{-3} $.

The code used for the results presented in this work is available upon request.

\section{Shear and flexion terms} \label{s:shear}
MSD-breaking effects for realistic ULDM cores, which have a finite radius and dynamical deviations from spherical symmetry due to time-varying interference patterns, generate shear and flexion terms that could, in principle, be sought after in lensing analyses unless they are smaller than or comparable to modeling uncertainties. Consider the lensing potential $ \Psi(\vec{\theta}) $, defined from the convergence as
\begin{equation} \label{conv}
\laplacian{\Psi} = 2 \kappa \ .
\end{equation}
For $ \theta \ll \theta_{\rm c} $, where $ \theta_{\rm c} \approx \lambda_{\rm dB}/(2D_{\mathrm{L}}) $ is an approximation for the core radius in angular variables, we can expand in $ \vec{y}:= \vec{\theta}/\theta_{\rm c} $,
\begin{align}
\begin{aligned} \label{psi_exp}
\Psi(\vec{y}) &\approx \Psi(0) + \pdv{\Psi(0)}{y_i} y_i+ \frac{1}{2} \pdv{^2\Psi(0)}{y_i\partial y_j} y_i y_j  \\
&+ \frac{1}{6} \pdv{^3\Psi(0)}{y_i\partial y_j \partial y_k} y_i y_j y_k + \ldots \ .
\end{aligned}
\end{align}
For systems at $ D_{\rm L} \approx \SI{1}{\giga\parsec} $ and the core radii of the ULDM we are considering, $ y \lesssim 0.2 $ in the region where images form.

The first term in the expansion is an unobservable constant, whereas the second term is degenerate with the unobservable source position. The lowest order term which has an impact on inference analysis is the third, which amounts to convergence and shear
\begin{align}
\begin{aligned}
\gamma_1(\vec{\theta}) &= \frac{1}{2\theta_{\rm c}^2}\qty( \pdv{\Psi(\vec{y})}{y_1 \partial y_1} -\pdv{\Psi(\vec{y})}{y_2 \partial y_2} ) \ , \\
\gamma_2(\vec{\theta}) &= \frac{1}{\theta^2_{\rm c}}\pdv{\Psi(\vec{y})}{y_1 \partial y_2} \ . 
\end{aligned}
\end{align}
Similarly, the third derivatives of $ \Psi $ correspond to flexion terms $ F, G $ (for definitions, see e.g.~\cite{Bacon:2005qr,Teodori:2023nrz}). The expansion in Eq.~\eqref{psi_exp} converges fast enough if (we remind that $ \Psi $ has the dimensions of a squared angle)
\begin{equation} \label{F_criterium}
\theta_{\rm c} \pdv{^3\Psi(0)}{\theta_i\partial \theta_j \partial \theta_k} \lesssim 1 \implies F, G \lesssim \frac{1}{\theta_{\rm c}} \ .  
\end{equation}

Our simulation outputs convergence on an equally-spaced grid, so it is natural to compute $ \Psi $ and its derivatives via FFT. 
From Eq.~\eqref{conv} (hatted quantities are in Fourier space),
\begin{equation}
	\hat\Psi_{\vec{k}} = - \frac{2}{k^2} \hat\kappa_{\vec{k}} \ .
\end{equation}
In a grid of length $ L $ with $ N $ points, with $ \vec{j} = (j_1,j_2) $, $ j_1,j_2 = 0,\ldots, N-1 $ labelling the grid point, 
we can write
\begin{align}
\begin{aligned}
	\Psi(\vec{\theta}) &= -\frac{2}{(2\pi)^2} \int \dd[2]{k} \frac{\e^{\iu \vec{k}\cdot\vec{\theta}}}{k^2} \hat\kappa_{\vec{k}} \\
	&\approx -\frac{2}{(2\pi)^2} \sum_{l_{1}, l_{2}=-N/2}^{N/2-1}  \e^{2\pi\iu \vec{l}\cdot\vec{j}/N} \frac{\hat\kappa_{\vec{k}}}{{l^2}}  \ , 
\end{aligned}
\end{align}
where we discretized the integral via $ \theta_i = j_i L/N $, $ k_i = 2\pi l_{i}/L $, $ l_i= 0,\ldots, N-1 $. 
With the discrete Fourier transform definitions\footnote{Most codes define the discrete Fourier transform with the sum starting from $ j,l=0 $ until $ N-1 $. Upon appropriate reordering exploiting the periodicity of the complex exponential, there is no difference.}
\begin{align}
\hat{\kappa}_{\vec{l}} &= \sum_{j_1,j_2=-N/2}^{N/2-1} \kappa_{\vec{j}} \e^{-2\pi \iu \vec{j}\cdot\vec{l}/N} =: F_N(\kappa)_{\vec{l}} \ , \\
\kappa_{\vec{j}} &= \frac{1}{N^2} \sum_{l_1,l_2=-N/2}^{N/2-1} \hat{\kappa}_{\vec{l}} \e^{2\pi \iu \vec{j}\cdot\vec{l}/N} := F^{-1}_N(\hat{\kappa})_{\vec{j}} \ , 
\end{align}
from which
\begin{equation}
\hat\kappa_{\vec{k}} = \qty(\frac{L}{N})^2 F_N(\kappa)_{\vec{k}} \ , 
\end{equation}
we can write
\begin{equation}
	\Psi = - 2\qty( \frac{L}{2\pi})^2  F^{-1}_N \qty(\frac{F_N(\kappa)_{\vec{l}} }{l^2 } ) \ . 
\end{equation}
Analogously, we can write, for $ \Psi $ derivatives,
\begin{align} 
	& \partial_i
	\Psi= -\iu \frac{L}{\pi} F^{-1}_N \qty(F_N(\kappa)_{\vec{l}}  \frac{l_i}{l^2 } ) \ , \\
	\label{shear_num}
	&\partial_i \partial_j
	\Psi = 2F^{-1}_N \qty(F_N(\kappa)_{\vec{l}}  \frac{l_{i} l_{j}}{{l}^2 } ) \ , \\  \label{F1_num}
	&\partial_i \partial_j \partial_m
	\Psi = \iu\frac{4\pi}{L} F^{-1}_N \qty(F_N(\kappa)_{\vec{l}}  \frac{l_{i}l_{m} l_{j}}{{l}^2 } ) \ .
\end{align} 

\begin{figure*}
	\centering
	\includegraphics[width=0.45\textwidth]{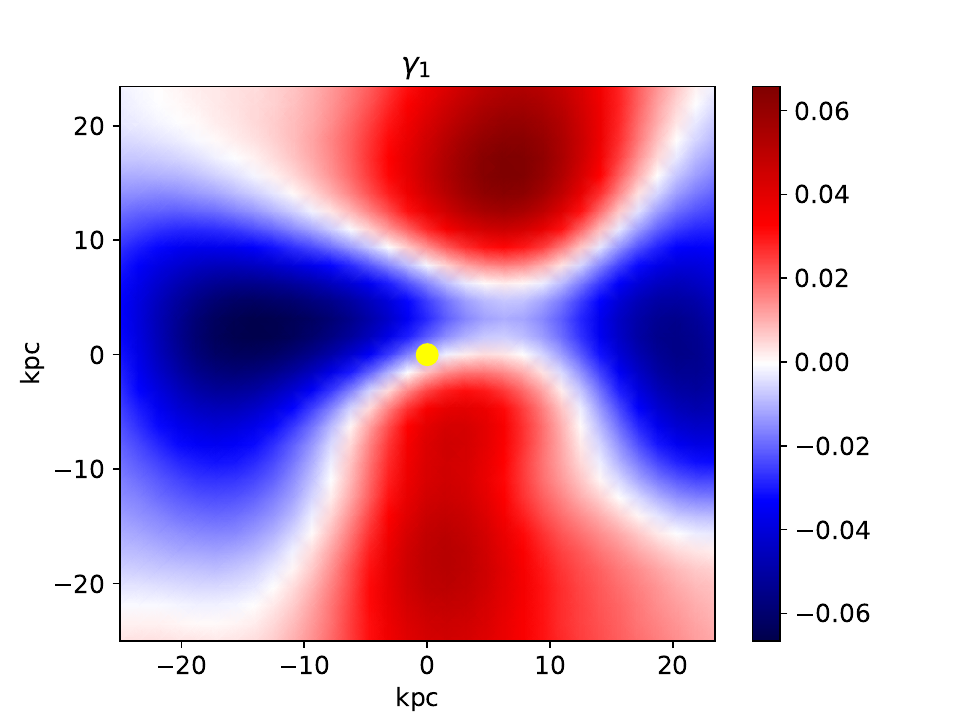}
	\includegraphics[width=0.45\textwidth]{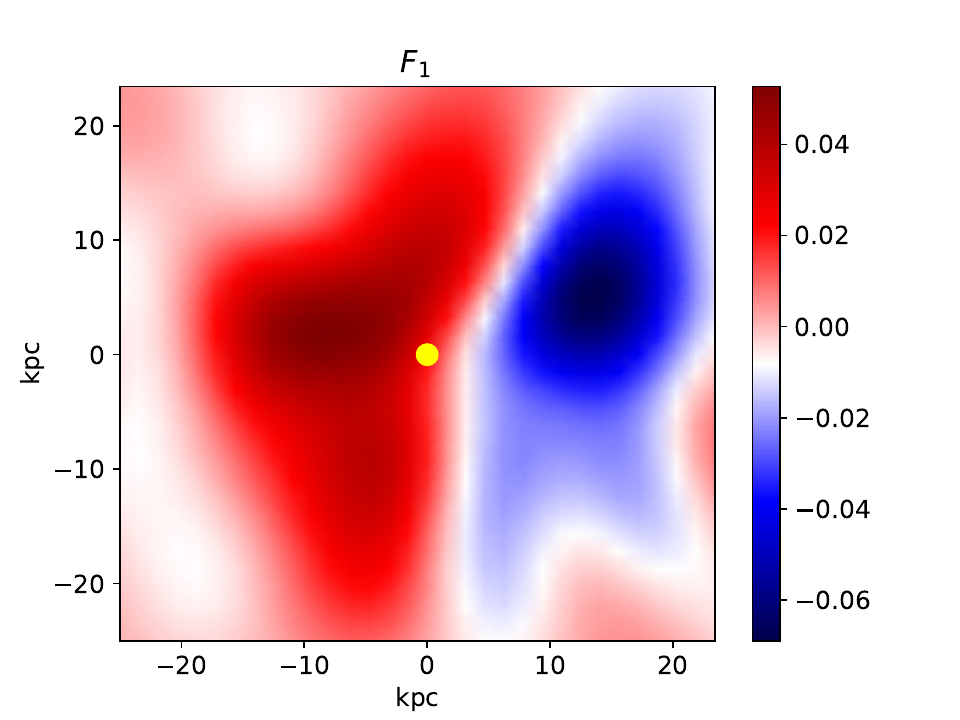}
	\caption{Plots of $ \gamma_1 $ and $ F_1 $ on the left and right respectively, referring to the convergence field in the bottom-left panel of Fig.~\ref{fig:12510e25}.}
	\label{fig:shear_F1}
\end{figure*}

In Fig~\ref{fig:shear_F1}, we show $ \gamma_1 $ and $ F_1 $ as proxy for shear and flexion magnitude, computed using Eqs.~\eqref{shear_num}-\eqref{F1_num}. These results refer to the system showed in the bottom-left panel of Fig.~\ref{fig:12510e25}, which has an high $ \Delta\kappa_{\rm c} \gtrsim 0.09 $. With a large $ \Delta\kappa_{\rm c} $, the system under consideration can yield possible detectable imaging errors. However, the magnitude of the shear near the center of the core is comparable to (and degenerate with) typical cosmological external shear, which is poorly constrained theoretically. Similarly, the flexion magnitude is not too far off from the flexion caused by a host group of the lens galaxy~\cite{Teodori:2023nrz} (notice that criterium Eq.~\eqref{F_criterium} is satisfied, justifying the relevance of the expansion in Eq.~\eqref{psi_exp}). This calls for the possibility that MSD-breaking effects from cores with $ \Delta\kappa_{\rm c} \gtrsim 0.03 $ could still be reabsorbed by other modeling parameters. Once the features of these ULDM cores are better constrained theoretically (e.g. with cosmological simulations), one can use mock systems to explore the extent of such degeneracies. 

\section{ULDM initialized with Burkert halo} \label{s:burk}
As a further validation, we here show some simulation results, where the initial ULDM halo is initialized as a Burkert halo, i.e. with target density profile
\begin{equation} \label{eq:burk}
\rho_{\rm Burk} (r) = \frac{\rho_{\rm s}}{(1+r/r_{\rm s}) (1+(r/r_{\rm s})^2) } \ .
\end{equation} 
We present the results in Tab.~\ref{tab:runs_Burk}. These results are consistent with those shown in Tab.~\ref{tab:runs}, although different initial conditions can lead to varying core formation time scales. This may result in underdeveloped inner cores compared to halos initialized with the NFW profile. This effect is particularly noticeable for the case where $ m = \SI{5e-25}{\electronvolt} $. This highlights the need for cosmological simulations to refine the theoretical predictions for the AxionH0graphy signal.

\begin{table*}
	\begin{tabular}{ccccccccccc}
		\toprule
			$ \frac{m}{10^{-25}\text{eV}} $ &  $ \frac{M_{200}}{10^{12} M_\odot} $& $\frac{\rho_{\rm sol}(0)}{10^3\si{M_\odot\per\parsec\cubed}}$ &$ \frac{M_{\rm sol}}{10^{11} M_\odot} $
		& $ \frac{r_{\rm c}}{\text{kpc}} $ & $ \frac{\lambda_{\rm dB}}{2\text{kpc}} $ & $\frac{r_{\rm E}}{\text{kpc}}$ & $10^2\delta_{\rm E}$ 
		& $ 10^2\Delta \kappa_{\rm c} $ & $  \delta_H\approx\kappa_{\rm c}(\theta_{\rm E})  $  & Conditions \\
		\midrule
$10$ & $16.9$ & $4.9^{+1.2}_{-0.6}$ & $0.1^{+0.0}_{-0.0}$ & $4^{+0}_{-0}$ & $4$ & $12.0^{+4.0}_{-4.0}$ & $1.2^{+0.5}_{-0.3}$ & $1.2^{+0.5}_{-0.8}$ & $0.03^{+0.00}_{-0.01}$ & 1 Burkert halo, ext NFW + Hernq. \\
\midrule
$5$ & $17.8$ & $4.0^{+0.6}_{-1.1}$ & $0.1^{+0.1}_{-0.0}$ & $7^{+3}_{-0}$ & $8$ & $12.6^{+4.2}_{-4.2}$ & $0.0^{+0.0}_{-0.0}$ & $2.2^{+1.1}_{-1.3}$ & $0.03^{+0.00}_{-0.01}$ & 1 Burkert halo, ext NFW + Hernq. \\
\midrule
$2$ & $16.9$ & $2.0^{+1.8}_{-0.3}$ & $1.1^{+0.4}_{-0.2}$ & $15^{+5}_{-0}$ & $21$ & $12.0^{+4.0}_{-4.0}$ & $1.6^{+0.7}_{-0.4}$ & $1.1^{+1.6}_{-1.1}$ & $0.05^{+0.01}_{-0.02}$ & 1 Burkert halo, ext NFW + Hernq. \\
\midrule
$1$ & $13.6$ & $0.6^{+0.0}_{-0.4}$ & $3.0^{+0.2}_{-1.3}$ & $37^{+0}_{-0}$ & $45$ & $8.4^{+3.1}_{-3.1}$ & $0.8^{+0.1}_{-0.3}$ & $0.2^{+0.1}_{-0.1}$ & $0.02^{+0.00}_{-0.01}$ & 1 Burkert halo, ext NFW + Hernq. \\	
	\bottomrule
	\end{tabular}
	\caption{Summary of simulation results for $\alpha_\chi=0.15$, when the ULDM halo is initialized as a Burkert halo, Eq.~\eqref{eq:burk}. Columns are the same as in Tab.~\ref{tab:runs}. }
	\label{tab:runs_Burk}
\end{table*}

\end{document}